# Semiconductor Thermistors

## Table of Contents

# Semiconductor Thermistors

Dan McCammon

Physics Department, University of Wisconsin, Madison, WI 53706 USA

**Abstract.** Semiconductor thermistors operating in the variable range hopping conduction regime have been used in thermal detectors of all kinds for more than fifty years. Their use in sensitive bolometers for infrared astronomy was a highly developed empirical art even before the basic physics of the conduction mechanism was understood. Today we are gradually obtaining a better understanding of these devices, and with improvements in fabrication technologies thermometers can now be designed and built with predictable characteristics. There are still surprises, however, and it is clear that the theory of their operation is not yet complete. In this chapter we give an overview of the basic operation of doped semiconductor thermometers, outline performance considerations, give references for empirical design and performance data, and discuss fabrication issues.

## 1 Introduction

Early thermal detectors were hampered by the lack of practical thermometers with good sensitivity at low temperatures. Thermocouples and metallic resistance thermometers rapidly lose sensitivity below room temperature. Semiconductor thermistors were in use at high temperatures, but the readily available semiconductors — mostly metal oxides — became much too resistive when cooled. All of the most sensitive devices discussed in R. Clark Jones' classic 1947 paper on the ultimate sensitivity of thermal detectors operated at room temperature [1].

By 1950 there were conference reports indicating that doped germanium could be a suitable thermometer at temperatures as low as 1 K [2], but reproducibility was elusive. Material purity and characterization were difficult issues, and providing low-noise electrical contacts was a poorly understood black art. By 1961, Frank Low had developed practical gallium-doped germanium devices operating at liquid helium temperature with noise equivalent power (NEP) below $10^{-12}$ W/Hz$^{1/2}$ [3]. The radio frequency resistivity of this material made it a reasonable match to free space, so the sensing element could also be used as an efficient radiation absorber, and these “germanium bolometers” were soon applied to the burgeoning field of infrared astronomy.

Today, developments by the semiconductor electronics industry have solved many of the materials and fabrication issues. Germanium and silicon are readily available with more than adequate purity, essentially noise-free contacts are easily formed by ion-implanted doping, and a host of techniques and machines are available that facilitate the fabrication of the thermistors, thermal isolation structures, and large arrays of detectors and their electrical interconnections. There have been considerable advances in theoretical understanding of the conduction mechanism. As we will see below, this understanding is far from complete, but empirical characterizations should be adequate for optimizing the design of low temperature detectors, and for predicting their ultimate performance with thermometers of this type.

## 2 Hopping Conduction

Shallow impurities in germanium and silicon have binding energies typically tens of milli-electron volts and are almost completely ionized at room temperature. As the temperature is lowered, the ionized fraction drops rapidly. This produces an increase in the resistance of lightly-doped material, but for most low temperature thermometry applications this is not the regime nor mechanism of interest. Instead, we are normally working at temperatures where thermal ionization of impurities is negligible. While the crystal structure is perfectly regular, the random substitution of dopant atoms at a small fraction of the lattice sites produces a disordered system that has, at sufficiently low temperatures, the electron transport properties of an amorphous material. The basic properties of such systems were determined by Anderson [4] and Mott [5]. They find that there is a critical doping density below which the conductivity goes to zero at zero temperature, and above which there is always a finite conductivity. We are interested in materials doped below this "metal-insulator transition", where charge transport takes place by phonon-assisted tunneling between impurity sites. The energy levels of these sites are effectively randomized by the long-range coulomb potential of charges distributed over distant sites, and the energy difference required in a given tunneling event or "hop" is made up by absorption or emission of a phonon of the required energy.

At sufficiently high temperatures, the distance barrier wins and hops take place to the nearest unoccupied site. This is referred to as "nearest neighbor hopping". At lower temperatures, the scarcity of high energy phonons favors longer hops as necessary to find an unoccupied site sufficiently close to the same energy that a phonon is more readily available. This is usually the regime of interest for low temperature thermometry, and the process is called "variable range hopping" (vrh). An extensive development of the theory can be found in the text by Shklovskii and Efros [6]. For vrh, the conduction is expected to behave as

$$R(T) = R_0 \exp\left(\frac{T_0}{T}\right)^p , \tag{1}$$

where Mott found $p = 1/4$ for the approximately constant density of states he expected at the fermi level. Efros and Shklovskii later showed that the changing coulomb interactions accompanying a hop should guarantee the existence of a parabolic gap in the density of states centered at the fermi level [7]. This modifies the result to make $p = 1/2$.

Measurements at the time, particularly on germanium and silicon, usually did not look much like this. Experiments gave a variety of $R(T)$ curves that were not reproducible from one sample to the next. However, data from samples doped by nuclear transmutation (NTD) or by careful ion implantation *do* show the expected behavior. This can be seen in Fig. 1, where the $T^{-1/2}$ behavior predicted by vrh with a "coulomb gap" is quite accurately followed over several orders of magnitude in resistance.

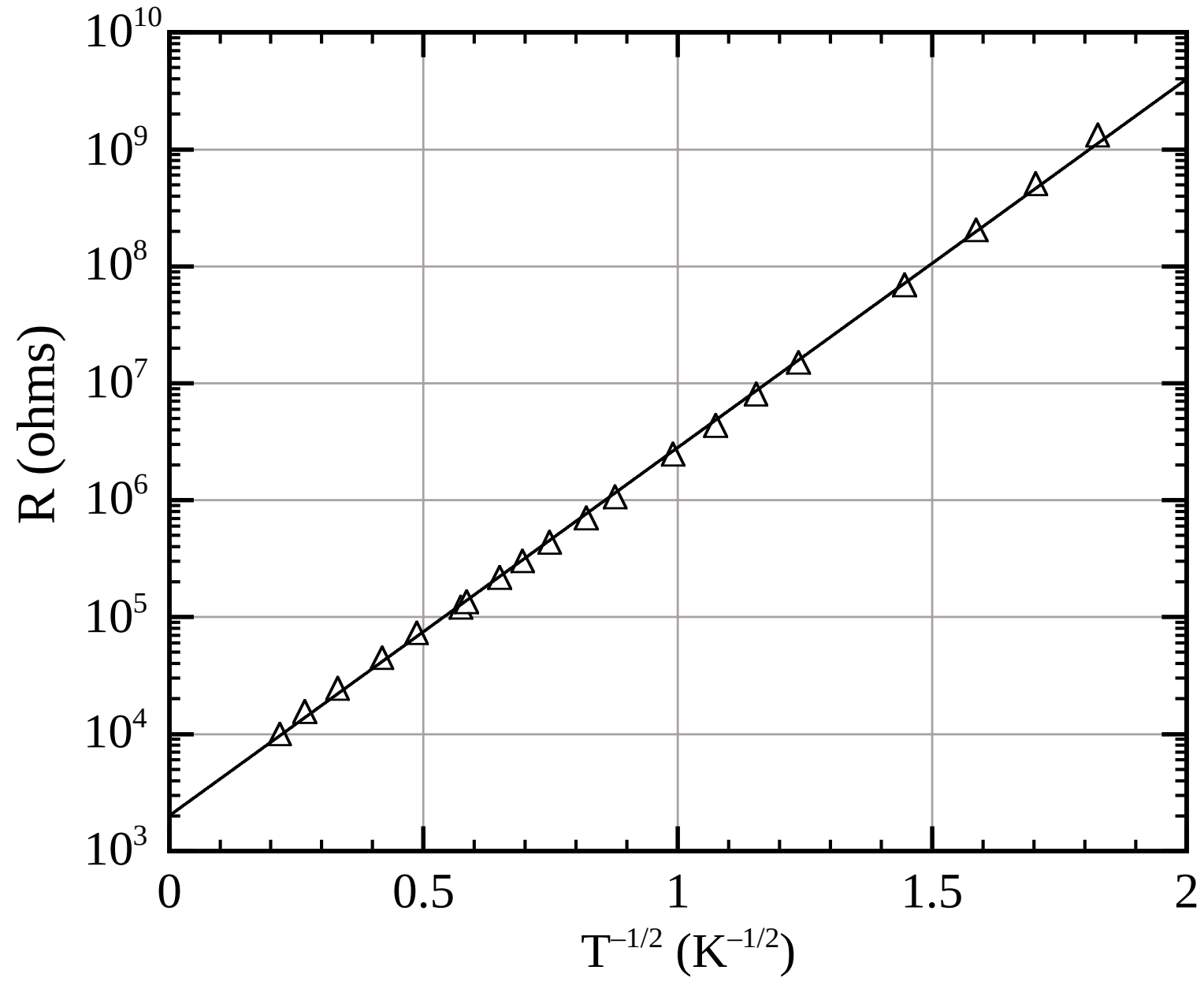

**Fig. 1.** Measured *R(T)* for ion-implanted silicon. The linear dependence of log(*R*) on $T^{-1/2}$ predicted by the variable range hopping model with a coulomb gap is observed over several orders of magnitude in resistance

Even modern melt-doped material seldom shows this clear coulomb gap behavior. The resistance usually flattens as the temperature is reduced, and the power-density effects discussed below become apparent at lower than expected power levels. Both of these observations are consistent with the existence of small-scale nonuniformities in the doping density, but it is not clear that this is the correct explanation.

## 2.1 Deviations from coulomb gap behavior

At sufficiently low temperatures, when $T_0/T$ > ~24, systematic deviations from this coulomb gap behavior are observed [8,9]. This is shown in Fig. 2, where *R(T)* is plotted for ion-implanted silicon samples with several different doping densities. It is clear that many of these curve upwards, but it is difficult to see a pattern in the deviations and in fact would be difficult to detect them at all had the lowest temperatures been included in the straight line fits. However, if we divide out the coulomb gap model fits, where $\log R \propto T^{-1/2}$, and plot the ratio as a function of the temperature normalized by $T_0$ for each sample as shown in Fig. 3, it can be seen that the deviations are quite systematic.

Shlimak observed similar behavior in arsenic-doped germanium, and suggested that a magnetic hard gap due to spin-spin interactions might be responsible [10]. There is evidence that *R(T)* reverts to coulomb gap behavior with applied magnetic fields > 1 T, which supports the idea that the deviations are some kind of magnetic effect [11].

From an experimental standpoint, this low-temperature rise can easily be masked by light leaks, RF pickup, or other extraneous heating effects, all of which tend to make measurements turn down below the intrinsic *R(T)* curve. It is useful to have an analytic expression that can be fit to measurements at higher temperatures, where these effects are

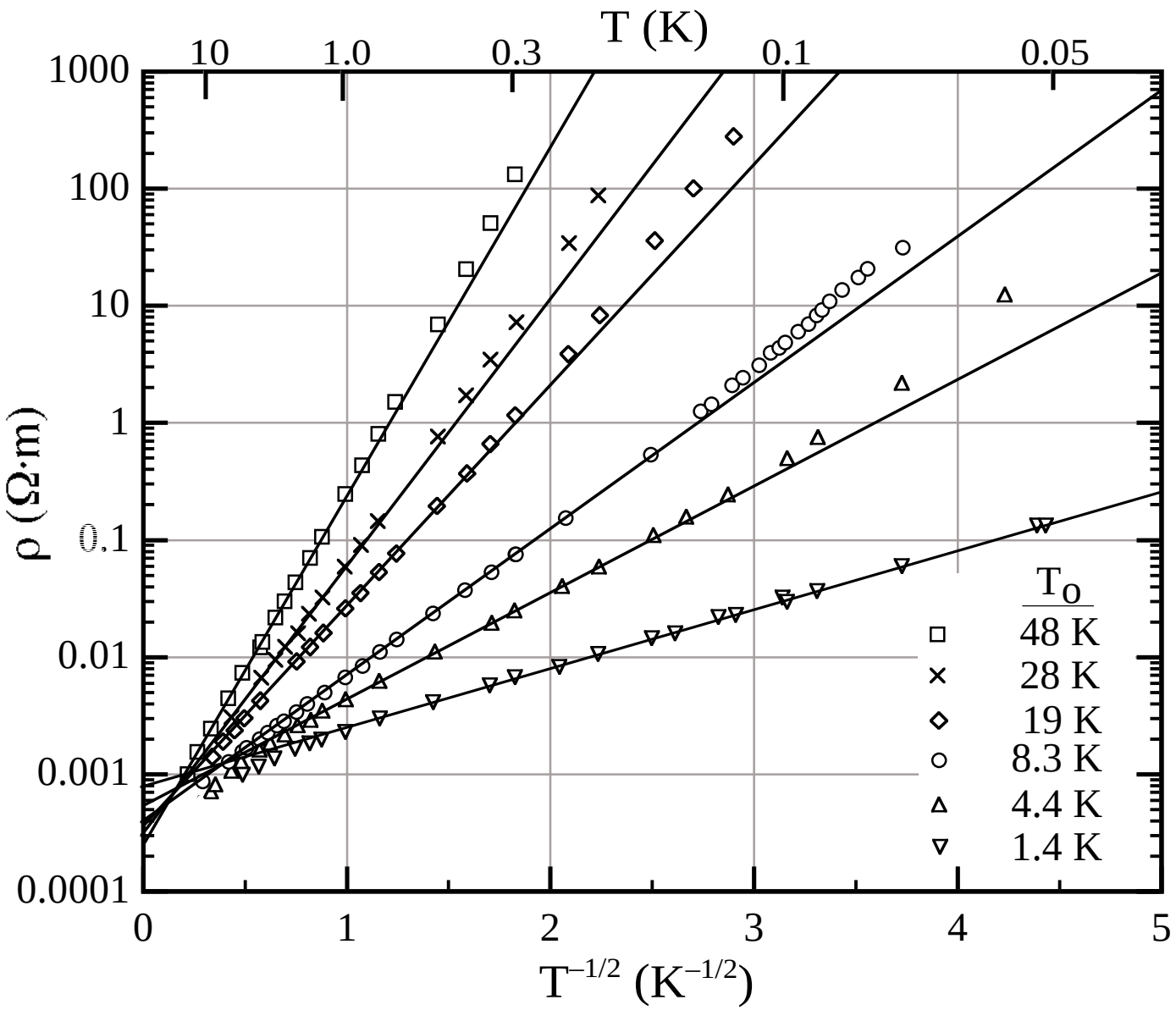

**Fig. 2.** Resistivity vs $T^{-1/2}$ for six ion-implanted Si:P,B samples. The different samples have different doping densities, which determines the value of $T_0$ in (1). The straight lines are fits to the coulomb gap model over a temperature range $6.5 < T_0/T < 24$. (from [8])

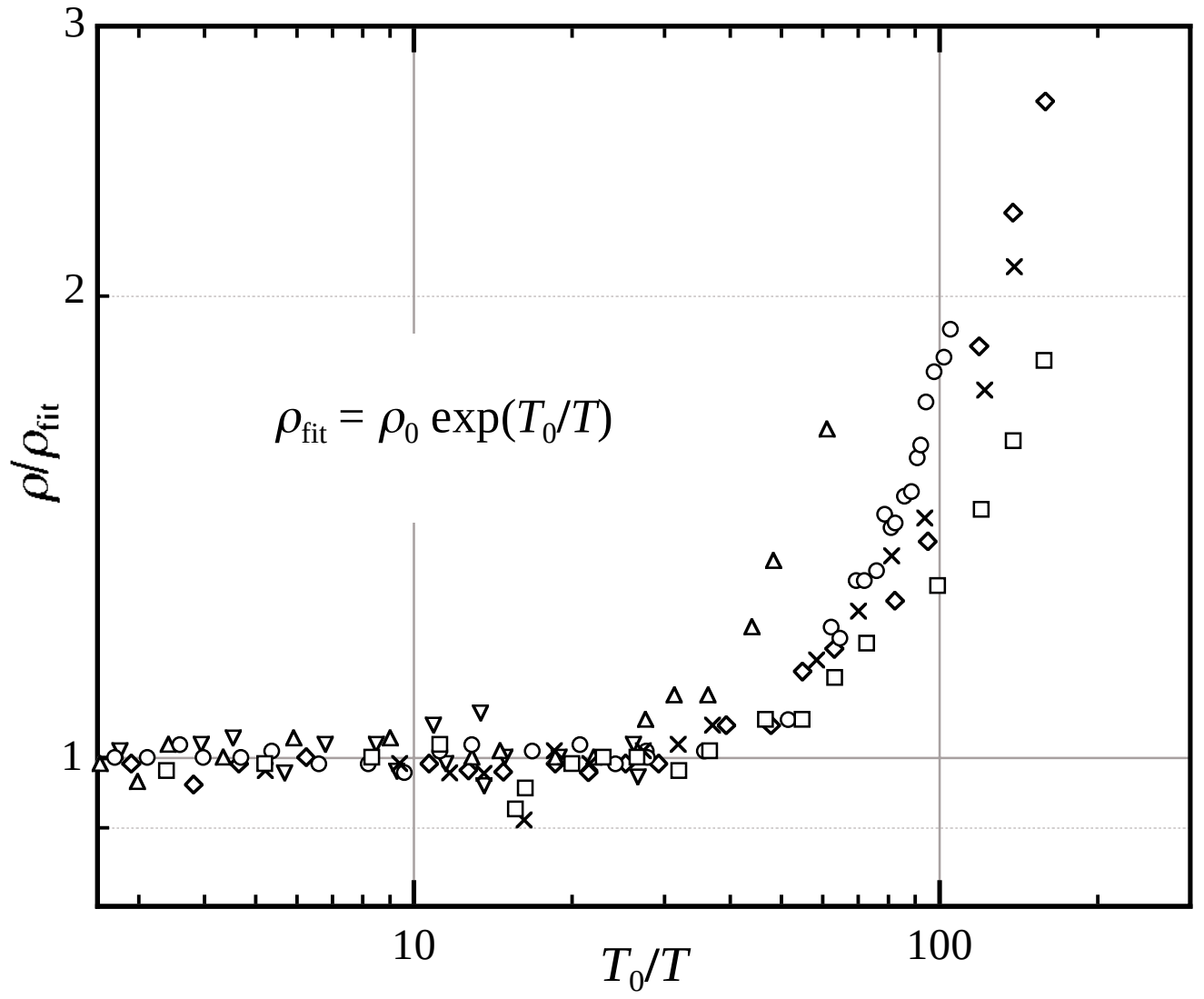

**Fig. 3.** Resistivity of samples in Fig. 2 divided by the best-fit coulomb gap model vs the normalized inverse temperature $T_0/T$. (from [8])

usually negligible, and extrapolated to compare with the lowest temperature measurements to determine the extent of any heating problems. Of course, it is also convenient for thermometric purposes to have such a function, since only a small number of calibration points are then required to fix the entire *R(T)* dependence. Wouter

Bergmann-Tiest has fit the deviations with a purely empirical function, which is reproduced here [12]:

$$R(T) = R_0 \exp\left(\frac{T_0}{T}\right)^{1/2} + R_0' \exp\left(\frac{T_0'}{T}\right)^{1/2}, \tag{2}$$

where $R_0' = R_0 \exp(2.522\, T_0^{-0.25} - 8.733)$ and $T_0' = 2.7148\, T_0 + 1.2328$.

This is not pretty, but it introduces no additional free parameters, and considerably increases the temperature range over which a good fit can be made. (In principle, there should be only one free parameter to the fit, which is $T_0$ and corresponds to the doping density. $R_0$ should be calculable from this and the sample geometry. In practice however this cannot be done with the required precision, so $R_0$ is normally fit simultaneously). The function is designed to give the same values for $R_0$ and $T_0$ as the coulomb gap model if the latter is fit only to the higher temperature data.

The generality of whatever effect is producing this low-temperature deviation is illustrated in Fig. 4, which shows *R(T)* for an NTD germanium sample, fit by both the simple coulomb gap model and by the "Wouter function" given above. Each has the same two free parameters The Wouter function was derived from the behavior of very thin (0.3 µm) ion-implanted silicon samples, but it provides a good fit for this relatively thick (200 µm) NTD germanium device.

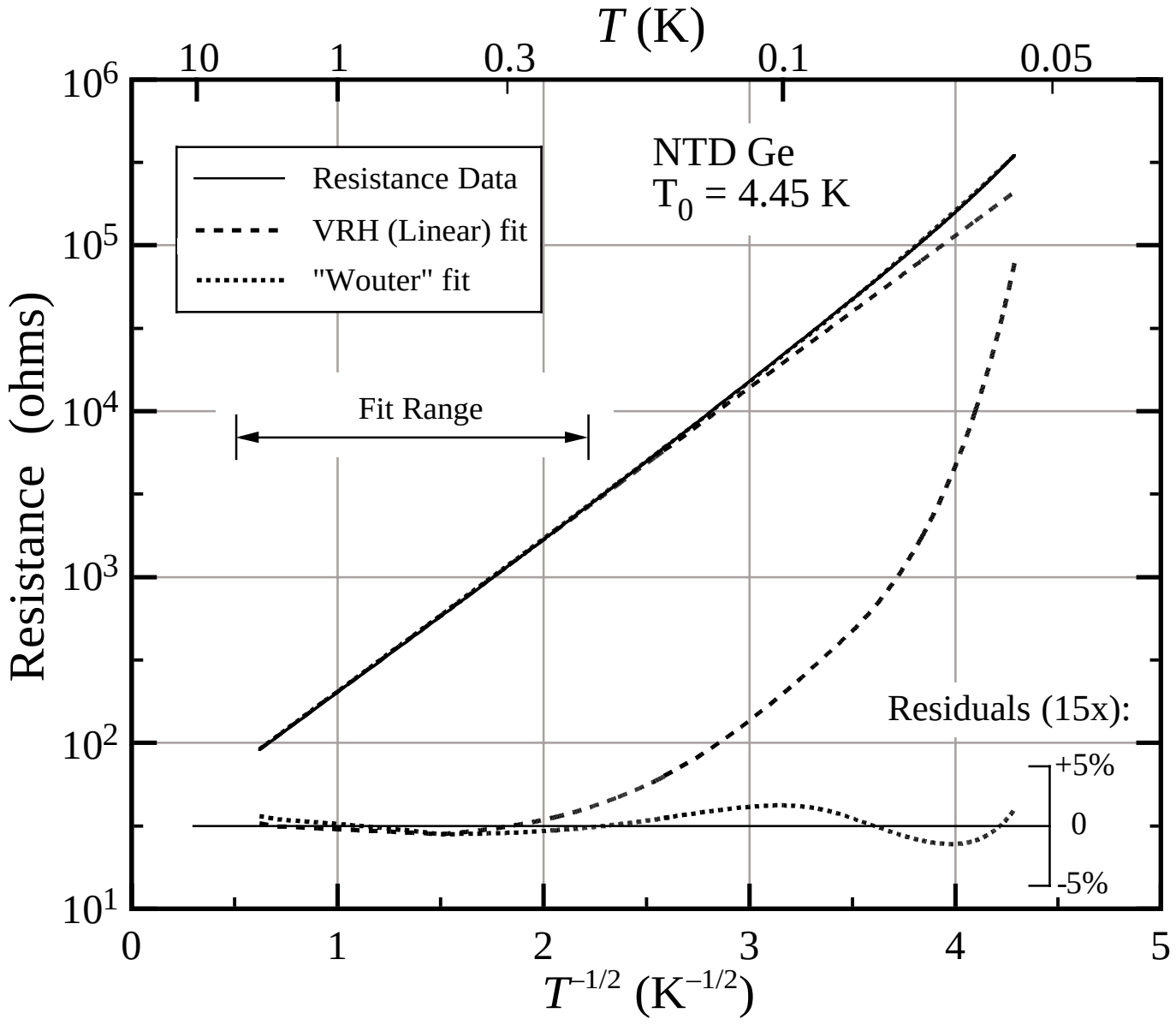

**Fig. 4.** Resistance of a 200 x 1000 x 2000 µm$^3$ NTD germanium thermometer, showing the best-fit coulomb gap model and a fit using the "Wouter function" given in (2). The fixed parameters of this function were derived from a set of thin ion-implanted silicon samples. Both functions were fit only to data for 4K < $T$ < 0.2K

This function works reasonably well, but its additive term offers little help with physical significance. Laura Rose Semo-Sharfman has made another fit to similar data, and chose to represent the correction as a multiplicative term [13]:

$$R(T) = R_0 \exp\left(\frac{T_0}{T}\right)^{1/2} \cdot \left(1 + A\left(\frac{T_0}{T}\right)^{B}\right). \tag{3}$$

where $A = (0.1884 - 0.01241 \ln T_0) \cdot 60^{-B}$ and $B = 2.074 + 3.179 \exp(-T_0/8.81)$. This actually doesn't fit quite as well as (2), but it is well within the systematic accuracy of the data. Woodcraft et al. [14] use a variable value of $p$ in (1) instead, but this gives poor fits to the data in [8].

## 2.2 Doping and device fabrication

This section contains quantitative information on doping germanium and silicon for device thermometers, and some examples of construction techniques.

***Neutron transmutation doped germanium.*** Extremely reproducible thermometers can be produced by irradiating germanium with reactor neutrons [15]. Natural germanium has four stable isotopes with substantial abundances, and the neutron cross sections and abundances of two of these conspire to produce gallium-doped (p-type) material that is 32% compensated with arsenic. Since the isotopes are chemically identical, they are presumably distributed perfectly randomly in the lattice, and the moderate neutron cross section makes the neutron flux uniform throughout even quite large blocks of germanium. These can then be cut up to make large numbers of very uniform thermometers. Plots of the resistivity $\rho(T)$ for a number of different neutron doses are shown in Fig. 5, and the fit $T_0$ vs net doping density is given in Fig. 6. The value of $\rho_0$ should also be a simple function of doping density, and a plot of this is given in [16]. However, the derived value is highly correlated with the $T_0$ fit, small systematic problems with the data make large differences, and there is little agreement in published values. Inspection of Fig. 5 shows that over a wide range of intermediate values of $T_0$, $\rho_0$ is within a factor of ~2 of 0.1 ohm cm, while samples with extreme values of $T_0$ can be a factor of 100 or more different.

The uniformity and predictability of NTD Ge thermometers is particularly valuable in the construction of large arrays of detectors. However, the penetrating power of the neutrons that helped provide the very uniform doping also means that it is impractical to mask the process and dope only selected areas in a crystal. So the thermometers must be cut to the optimum dimensions, and then individually attached to the detector elements. This is not a drawback, however, when the elements are very large and individually mounted, as in the CUORE project [18]. For arrays of small detectors, clever use of hybrid circuit mounting techniques that treat the thermistors as components to be "bump-bonded" can allow the use of integrated wiring and somewhat automated assembly [19].

***Ion-implanted silicon.*** Silicon requires very large neutron doses to dope by transmutation, and this is seldom attempted. However, doping by implanting ions from a beam with kinetic energies from tens of keV to a few MeV is a well-developed technique in the semiconductor electronics industry. This allows penetration depths of up to ~1 µm,

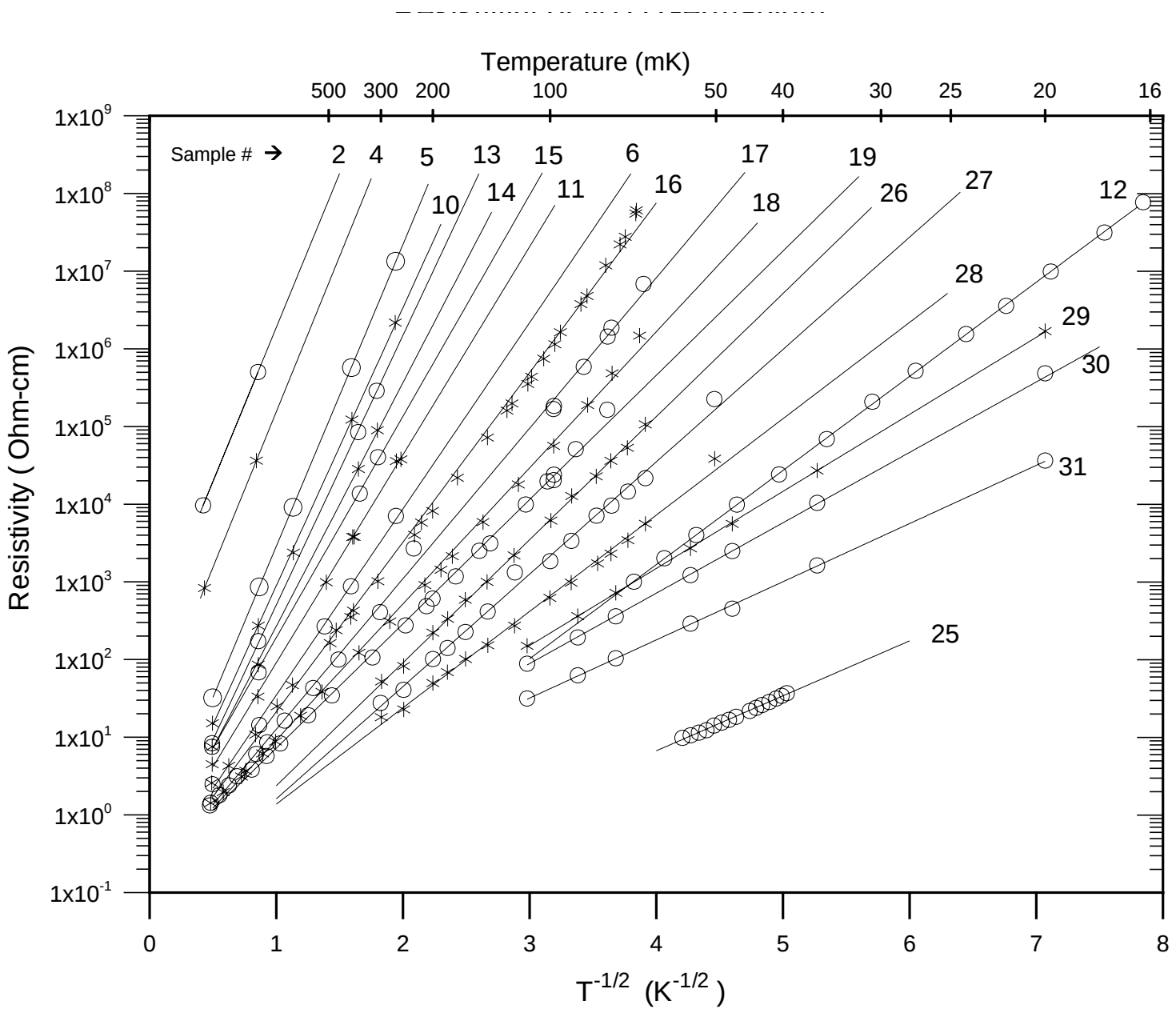

**Fig. 5.** Resistivity vs temperature for a number of samples of transmutation doped germanium exposed to different neutron fluences [17]

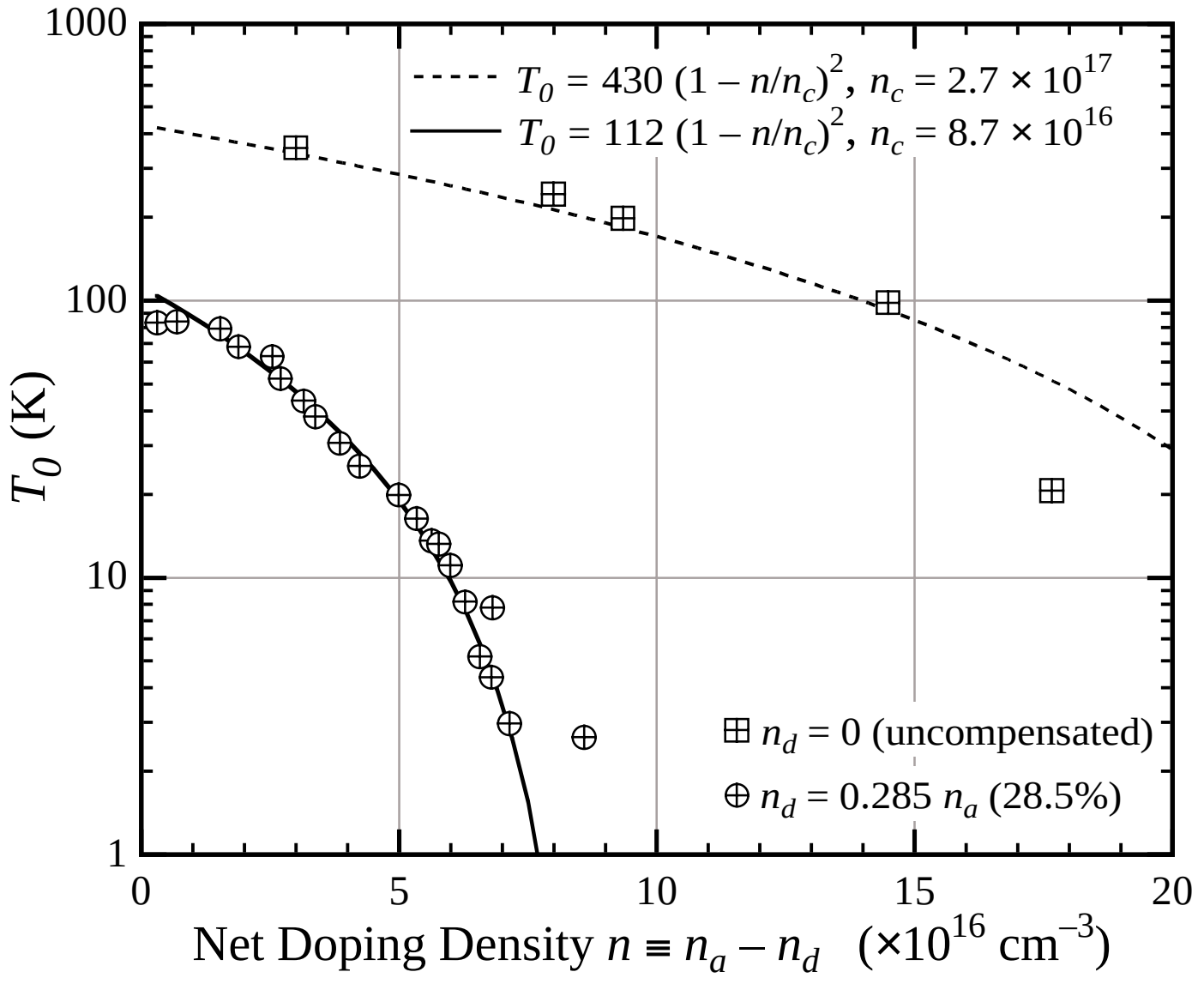

**Fig. 6.** Doping density parameter $T_0$ vs net doping density $(n_{\text{acceptor}} - n_{\text{donor}})$ for neutron transmutation doped germanium. Data provided by [17]

but results in an approximately Gaussian density profile with depth. Uniform densities can be obtained by superimposing implants with several different energies and carefully designed doses to produce a flat-top profile [8,9], or by implanting a single dose of each ion into a thin piece of silicon, preferably capped with $SiO_2$ on both sides, and then treating it at high temperature to allow the implanted ions to diffuse completely and uniformly throughout the thickness [20]. The diffusion times and temperatures are practical up to a thickness of a few microns.

The great advantage of this doping method is that it can be masked by standard photolithographic techniques, allowing the simultaneous fabrication of large numbers of thermistors than can have almost arbitrarily small dimensions, and have fully integrated electrical connections [21,22]. Silicon also has excellent mechanical and thermal properties, and many techniques exist for fabricating mechanical structures from it that can be used, for instance, for thermal isolation of the individual detector elements. Fig. 7 shows a sketch of the structure for a single pixel of an X-ray detector array, where a limited region is doped to form the thermometer and undoped silicon is used for the mechanical structure, including the thermal isolation beam supports. Fig. 8 shows a photograph of the pixel, and of the entire 6x6 array after the X-ray absorbers have been attached. Techniques exist that should make it possible to fabricate the absorbers monolithically, rather than attaching them by hand, but these have not yet been perfected.

Ion implanted thermistors of the stacked-implant variety have been plagued by a lack of reproducibility that is not understood. Both doses and energies can be measured with more than adequate precision, but the run-to-run repeatability is poor enough that it is common to implant a series of wafers with slightly different doses, and then pick the one that comes closest to the desired resistivity. Fortunately, uniformity across a wafer is generally good, and the steps in implant density within a single processing run give a monotonic sequence. Fig. 9 shows $T_0$ vs doping density for silicon, but it should not be taken too seriously, as the scatter among the data points shows. Diffused thermistors seem more repeatable, although there is not much experience with them yet.

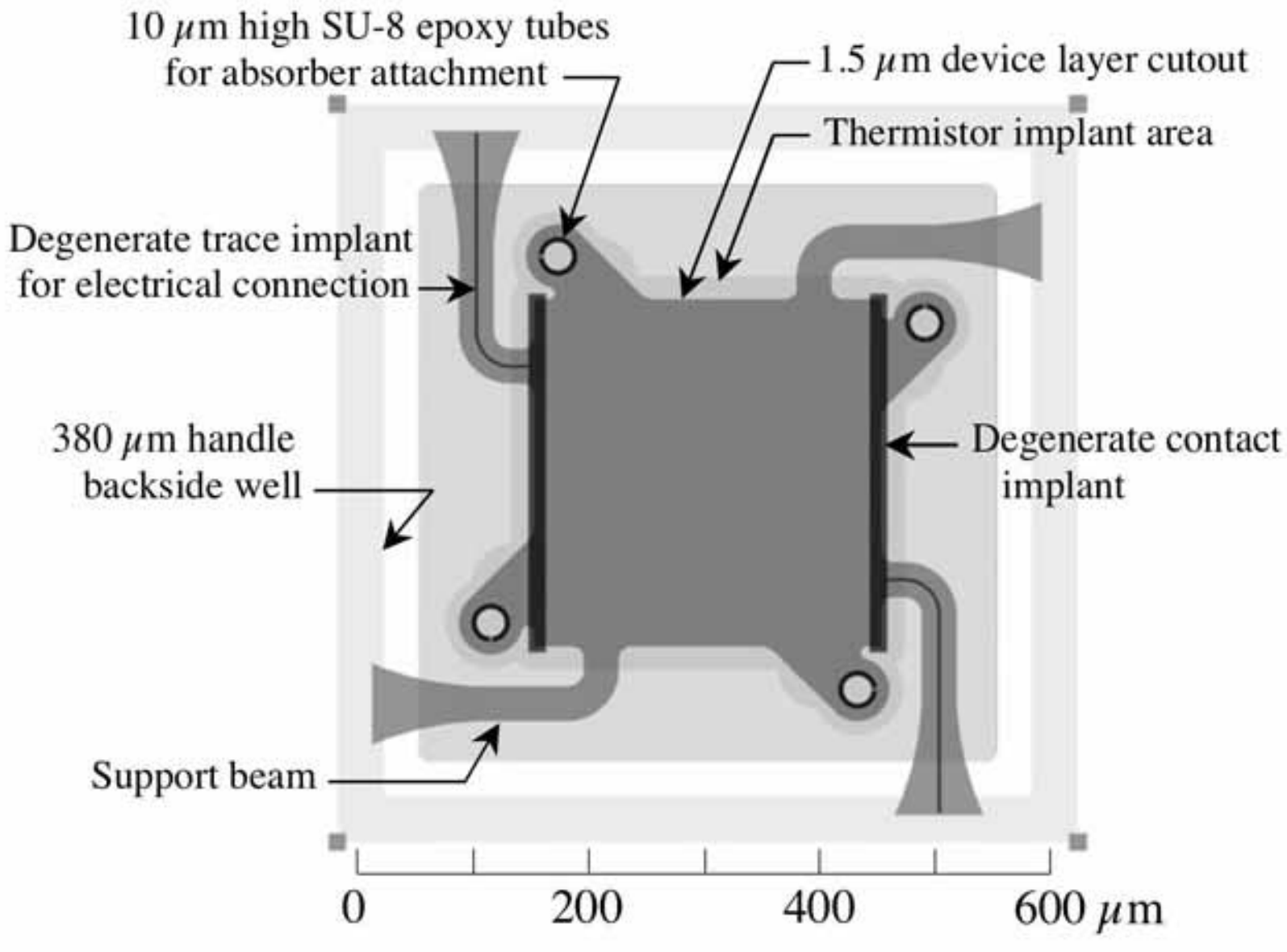

**Fig. 7.** Structure of one pixel of a monolithic X-ray detector array with ion-implanted thermistors.

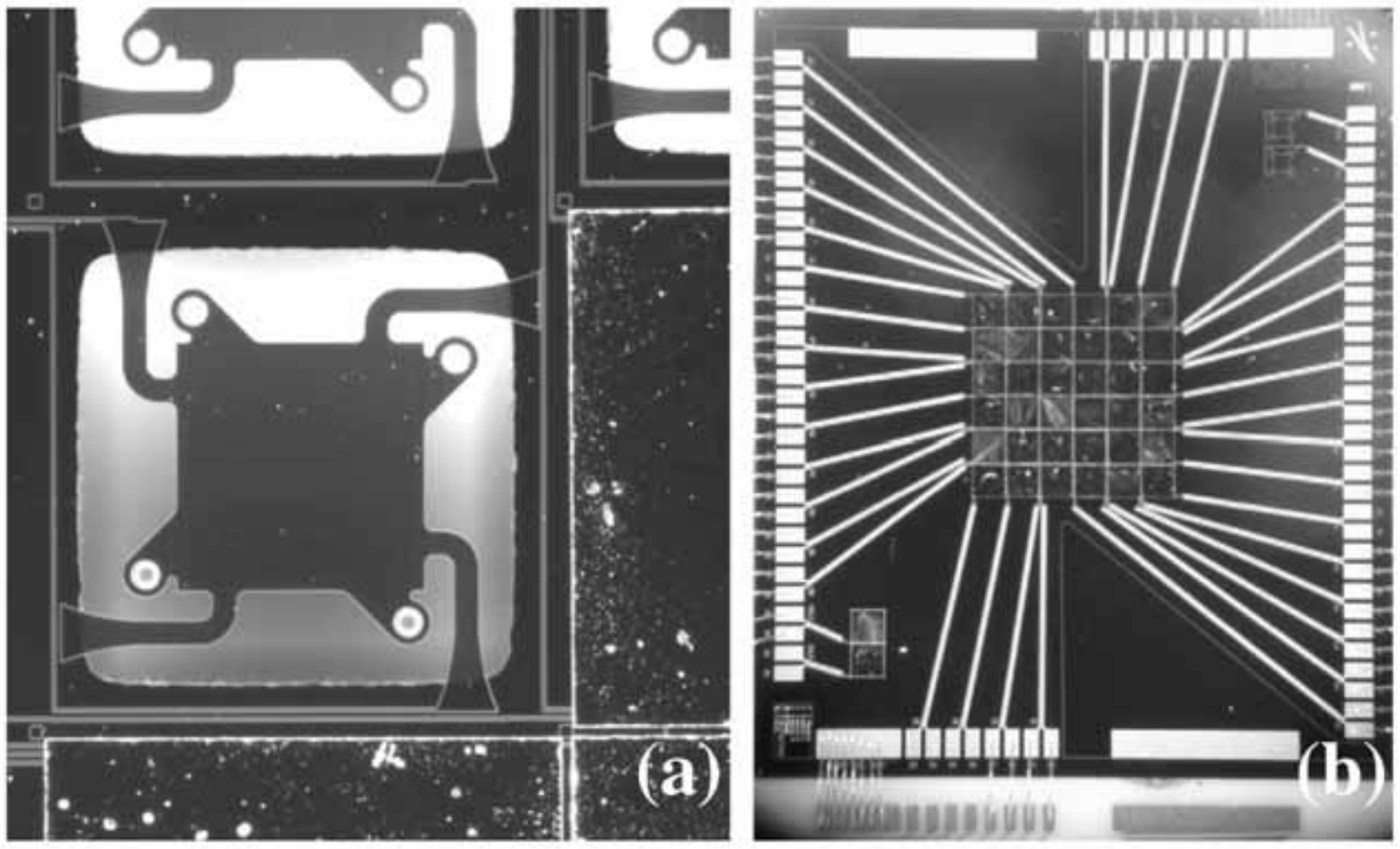

**Fig. 8. a)** One pixel of a monolithic silicon array for the XRS instrument on Astro-E2. The small cylinders on the short curved arms are attach points for the HgTe X-ray absorbers. **b)** Full view of the 6x6 array after the absorbers are attached with epoxy

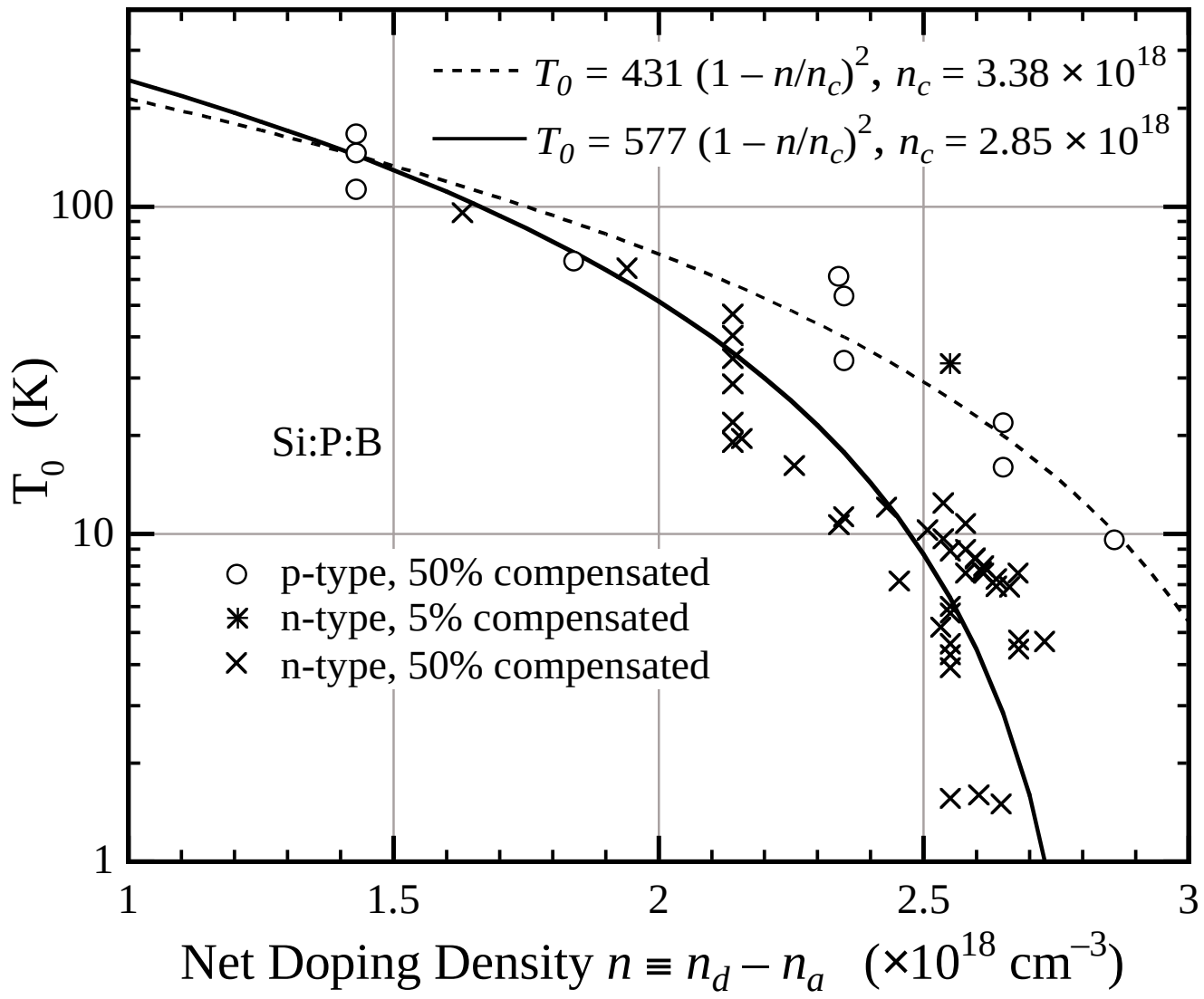

**Fig. 9.** $T_0$ vs net doping density for ion-implanted silicon. Most data are from [8], but the estimated thickness of these stacked implants has been normalized to results from the smaller number of diffused samples with accurately determined thicknesses, resulting in a 60% increase in effective thickness and a corresponding reduction in the derived doping densities

Reference [8] has a figure showing $\rho_0$ as a function of $T_0$ for implanted silicon. The difficulties are the same as for germanium, and this should be regarded as a rough guide only. The variation seems even smaller than for germanium, however, and $\rho_0 \approx$ 0.05 ohm cm over a wide range of $T_0$.

# 3 Electrical Nonlinearities

It might seem that doped semiconductors should make almost ideal thermometers. The figure of merit $\alpha \equiv d\log R/d\log T$ from Ch. 1 is just $\alpha = 0.5\,(T_0/T)^{1/2}$ for coulomb gap *R(T)*. One can make $T_0$ arbitrarily high by doping lightly, so it should be possible to make $\alpha$ as large as desired. At the same time, it is easy to fabricate ion-implanted sensors with such small volumes that the heat capacity contribution of the thermometer could be negligible, despite the relatively high specific heat of the doped material.

However, the effects described in this section introduce severe and in most cases fundamental limits to the extent that small volume and high sensitivity can be pursued. None of them is entirely understood theoretically, but empirical data are available (at least in principle) that allow optimum values for thermistor size and $T_0$ to be determined for a given application. These effects also introduce intrinsic limits to the speed of semiconductor thermometers.

## 3.1 Electric field effects

Phonon-assisted tunneling is an inherently non-linear process, and is expected to appear linear only in the limit of small electric fields. There are several models that differ in detail, but most can be approximated by

$$R(T,E) = R(T,0)\exp\left(-C\frac{eE\lambda}{kT}\right), \qquad (4)$$

where $E$ is the electric field, $R(T,0)$ is the resistance in the limit of low fields — the coulomb gap function in this case — $C$ is a constant of order unity, $e$ is the electronic charge, and $\lambda$ is a characteristic hopping length that in most models scales as $T^{-1/2}$ [23]. See also discussion and references in [24]. This type of behavior is observed in doped germanium and silicon under certain conditions [24,25,26], as shown in Fig. 10.

In the linear theory of Ch. 1, this behavior is represented by a local slope $\beta \equiv \partial R/\partial V\big)_T$ evaluated at the operating point. This is negative, and always acts to reduce thermometer sensitivity. Raising $T_0$ by decreasing the doping concentration makes $\lambda$ larger and increases the magnitude of this term. To optimize thermometer design, it would be nice to know $\lambda$ as a function of $T_0$ and $T$, particularly for NTD Ge, as we will see below. References [24, 25, 26] and [27] have modest amounts of data for Ge, and less modest amounts of disagreement. This is clearly an area where there is a shortage of needed engineering data.

## 3.2 "Hot electron" effects

As can be seen in Fig. 11, the standard electric field effect form given in (4) does not always describe the observed behavior very well. Over most of the parameter range of interest for low temperature detectors, field-dependence can be better fit by an analog of the hot electron effect in metals [24,28,29]. If the bias power is dissipated in the conduction electron system and sunk through the crystal lattice, one can envision an effective thermal conductivity between the electrons and phonons. Again by analogy to other thermal conductivities, this is assumed to have the form $G_0 T^{\beta}$, making the power transfer

$$P = \frac{G_0}{\beta + 1}\left(T_{\mathrm{e}}^{\beta+1} - T_{\mathrm{lattice}}^{\beta+1}\right). \tag{5}$$

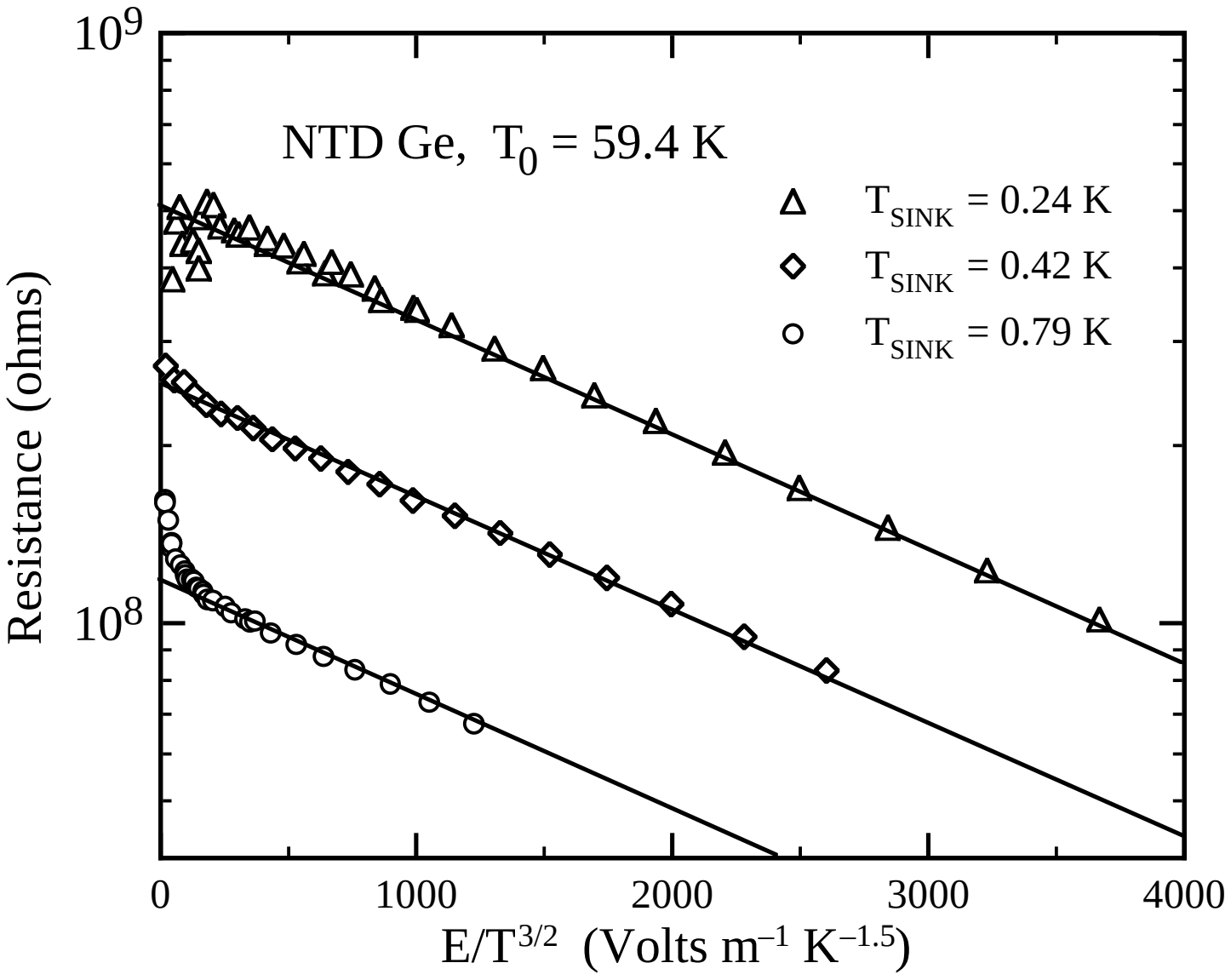

**Fig. 10.** Effect of electric field on resistance for a NTD Ge sample with $T_0$ = 59.4 K at three different heat sink temperatures. The sample was fixed to the sink with a high thermal conductivity. According to (4) with $\lambda \propto T^{-1/2}$, the data should follow parallel straight lines. (from [24])

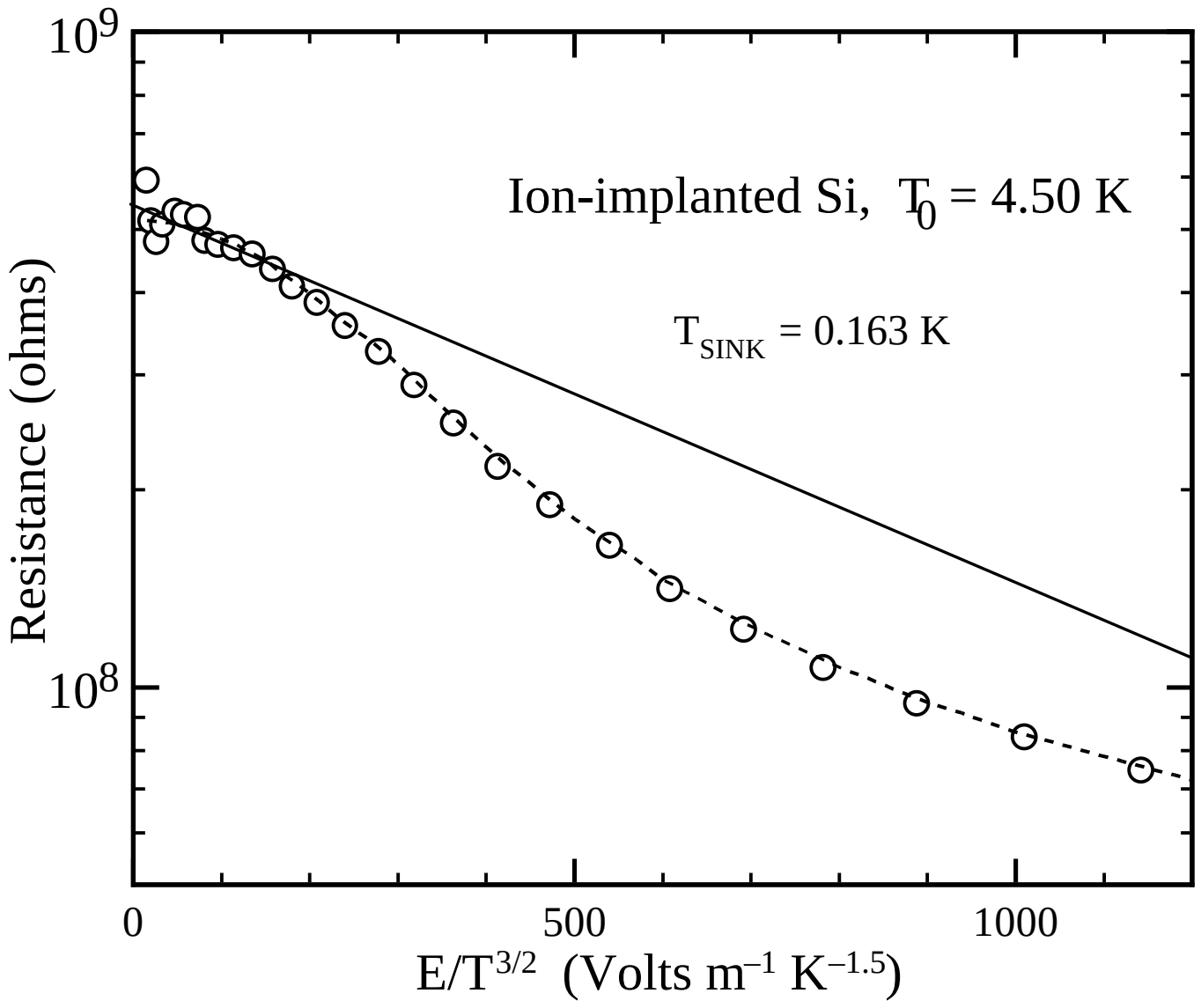

**Fig. 11.** As in Fig. 10, but for an ion-implanted Si sample with $T_0$ = 4.5 K , operated at 0.163 K. The solid line is a field-effect model. The dashed line is the hot electron model described in the text

In such a model, one further assumes that the resistance is a function of $T_{\mathrm{e}}$ only. (Note that this $\beta$ is unrelated to one used for thermistor voltage dependence.) The *R(T)* function can be calibrated in the limit of small bias power, where $T_{\mathrm{e}} \approx T_{\mathrm{lattice}}$. Then $G_0$ and $\beta$ are fit to resistance data taken with a range of bias powers. Figure 12 shows that this form fits well over a wide range of electron and lattice temperatures, although $\beta$ is generally found to be ~5, rather than the 4 expected and measured for metals.

Zhang et al [24] investigated a variety of ion-implanted Si and NTD Ge thermistors with a wide range of doping densities. They found that the difference between the behaviors of Figs. 10 and 12 is not that one is Ge and the other Si, but is rather the combination of operating temperature and doping density. Devices with high $T_0$'s operated at low temperatures had behavior well described by the field effect model of (4), while low $T_0$'s and high operating temperatures gave results better represented by the hot electron model of (5). Thermistors operating near the borderline, roughly described by $T_0/T \approx 100$, didn't fit either model well.

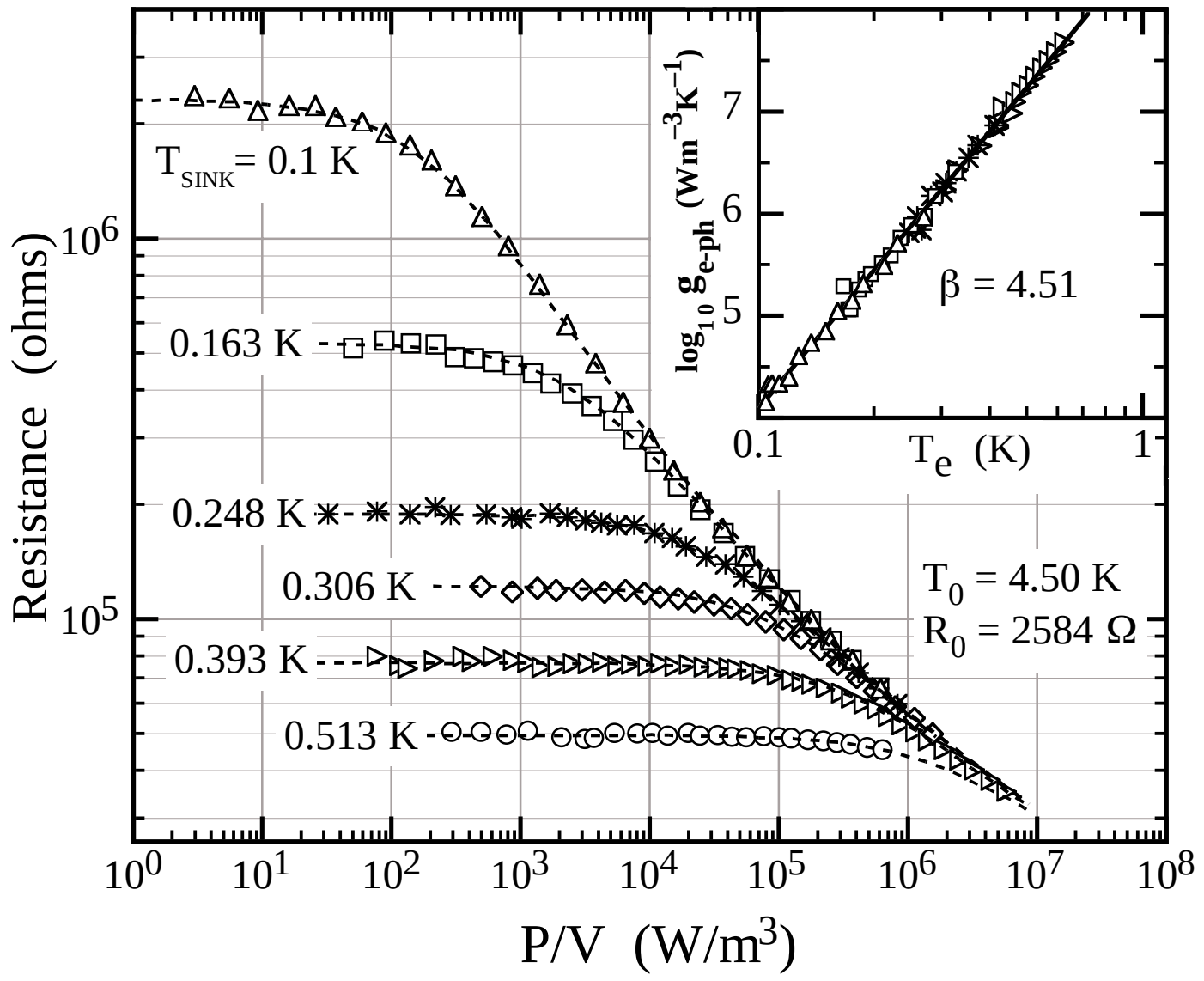

**Fig. 12.** Resistance vs bias power for an ion-implanted Si device. Data series are labeled by lattice temperature. The lines show the hot electron model, with electron temperature derived from (5) and resistance from this $T_{\mathrm{e}}$ and (2). The parameters $T_0$ and $R_0$ are fit to measured values of $R$ in the limit of low power. All data series are fit simultaneously with the same values of $G_0$ and $\beta$. The inset plots $G_{\text{e-ph}} \equiv d(P/V)/dT_{\mathrm{e}}$ vs $T_{\mathrm{e}}$, showing a single power-law dependence over four orders of magnitude in $G$. (from [24])

They suggested that both effects were always operative, with one or the other dominating in most parts of $T_0$-$T$ space, but with both required to explain the behavior in the boundary region, as evidenced by the data shown in Fig. 13 for an Si detector. Piat et al [27] found it essential to use a combined model to fit high quality data on NTD Ge thermometers intended for temperature control on the Planck-HFI instrument. The combined model simply substitutes $T_e$ from (5) for $T$ in the field effect model (4). The mean hopping length $\lambda$ is assumed to scale as $T^{-1/2}$, which is observed to be an approximation at best [24, 25, 26] (and [26] finds ~$T^{-1}$), but is adequate for fitting over a limited temperature range.

Like the field effect, hot electron behavior is an important limitation on detector performance, and its parameters are needed to optimize detector design. Zhang et al [24] characterized a large number of ion-implanted Si devices with a wide range of doping densities. They used the hot electron model to determine values for the electron-phonon thermal conductivity parameters $G_0$ and $\beta$ in (5) as a function of doping density, as shown in Fig. 14. Zhang et al were measuring thin stacked-implant devices, where the effective thickness is somewhat ambiguous. We have used measurements on a small number of diffused implant thermometers, where the thickness is precisely known, to normalize the volumes used for computing $G_0$/V in Fig. 14b. The devices used for normalization had $T_0$ near 7 K, and the correction was about 40%. It is possible that the correction should really be a function of $T_0$. Also, no corrections were made for electric field effect, which could be a significant contributor at the higher $T_0$'s. This would tend to reduce $\beta$ and increase $G_0$ in this region.

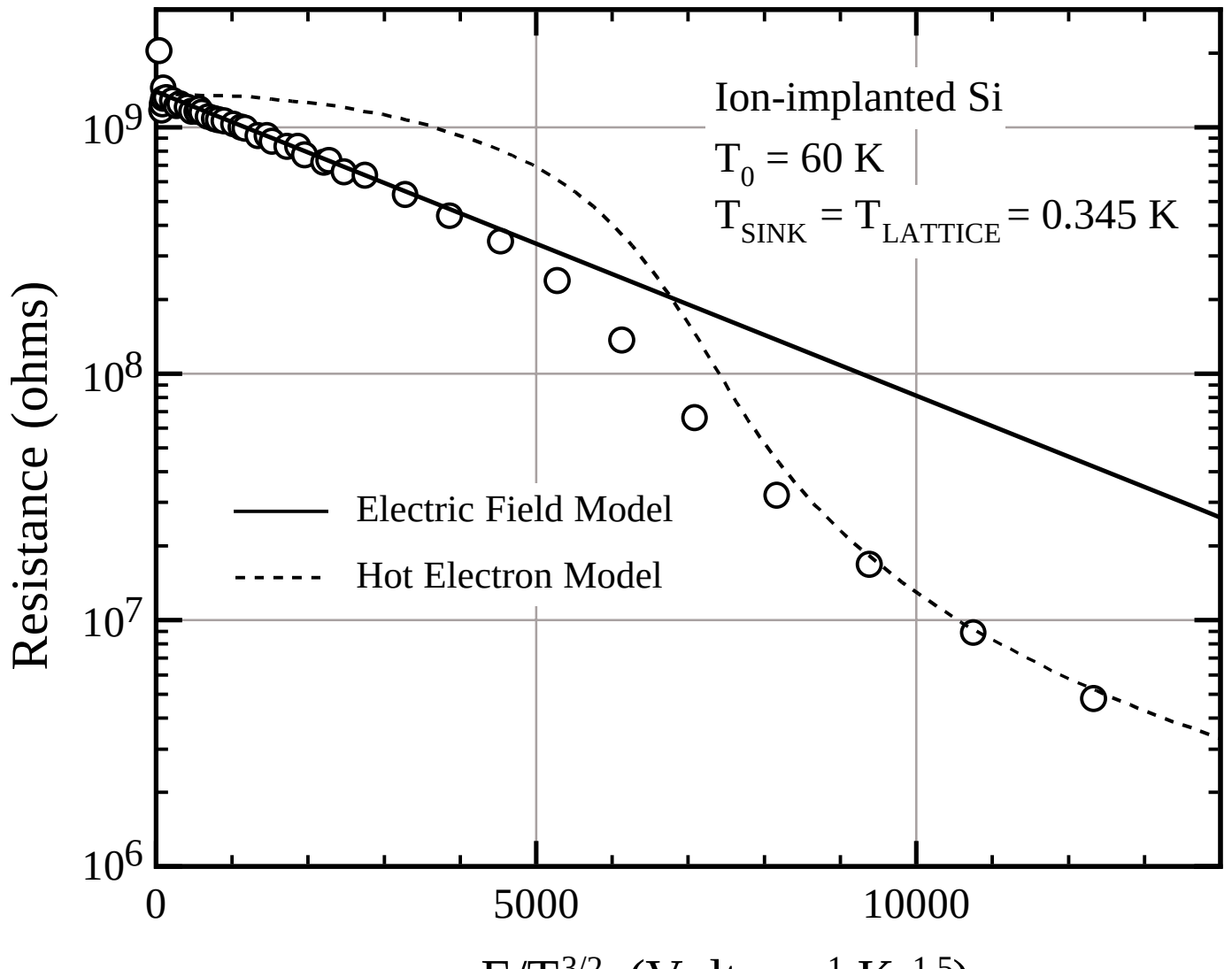

**Fig. 13.** Data for an ion-implanted Si sample operated near the $T_0$–$T$ borderline that separates good fits to the field effect and hot electron models. The solid line shows an electric field effect model (4), while the dashed line is for a hot electron model

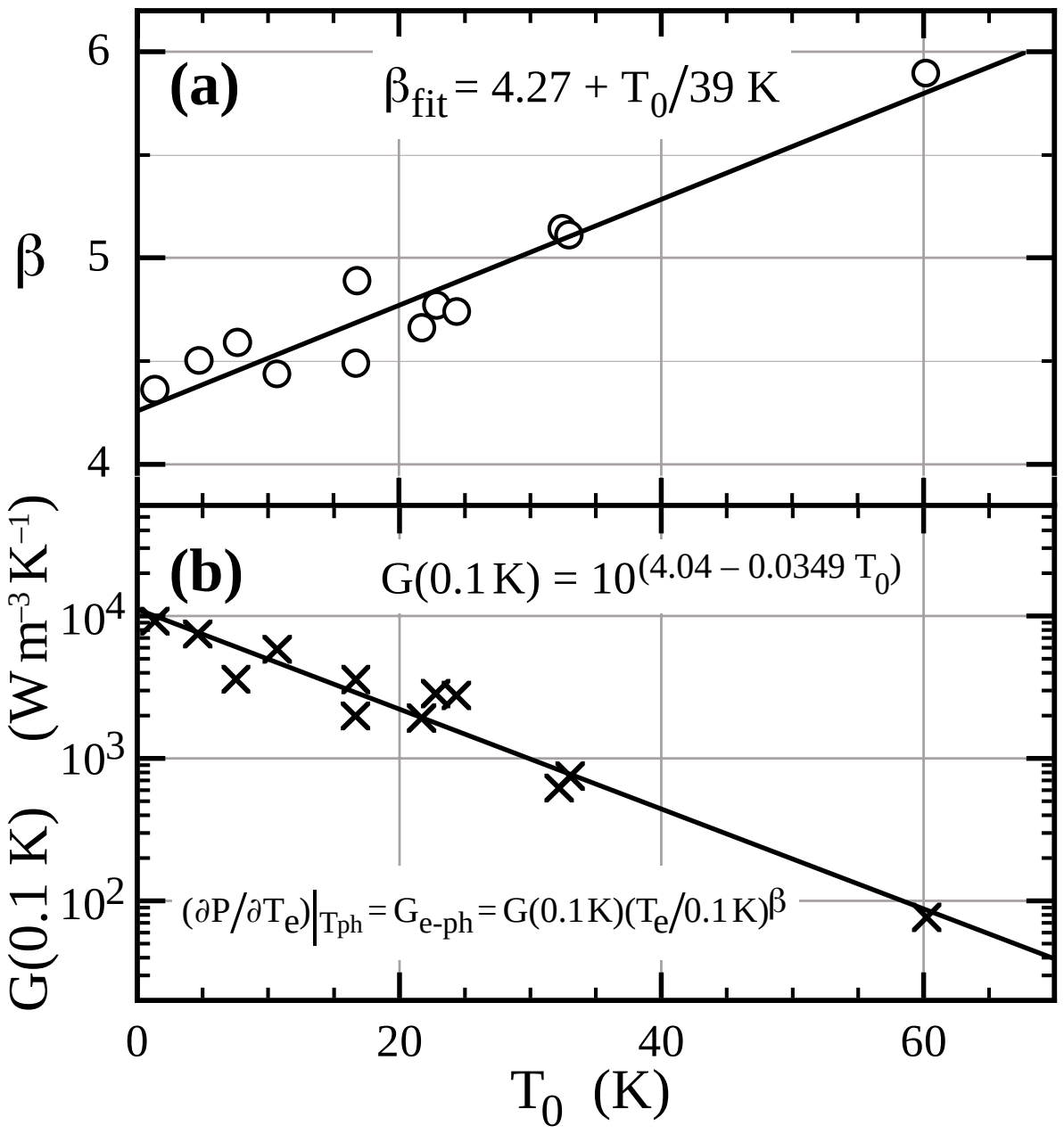

**Fig. 14.** Parameters fit to the hot electron model for a number of ion-implanted Si devices. **a)** Best-fit power law exponent $\beta$ as a function of $T_0$. **b)** $G_{\text{e-ph}}$ at 0.1 K as a function of $T_0$

While it is relatively easy to ensure that the lattice temperature remains constant as bias power is increased with the implanted Si samples, for practical reasons related to the generally quite different sample geometry and mounting, this is more difficult for NTD Ge. Again because of differences in the usual thermometer geometry, NTD Ge seldom has useful resistance values where hot electron effects are entirely dominant. Partly due to these factors, there is less quantitative data available on $G_0$ and $\beta$ for Ge. References [27, 28] and [29] have modest amounts of data on this, and again it is not in good agreement. Piat et al [26] were the only ones to make a simultaneous fit of the hot electron and electric field effects, so their results are probably the most reliable.

The difficulty with this "hot electron" model is that it has no basis in current theory of variable range hopping conduction. The doping of these devices is far on the insulating side of the critical density, and the electrons should be strongly localized. The electrons can change their energy distribution by tunneling from site to site, with emission or absorption of phonons making up the energy difference. But for the electrons to set up a thermal distribution of their own independent of the phonons would require that their energy be somehow delocalized while the electrons themselves are not [30]. It is interesting that $\beta = 5$ is expected for three-dimensional phonons in semiconductors doped on the metallic side of the metal-insulator transition, although this appears to have been verified only for the two-dimensional case, where $\beta = 4$ [31]. It is also true that the thin implants measured in [24] should be close to two-dimensional.

Since a real electron heating effect would depend on power density, while normal electric field effects depend on E, it might at first seem that we could distinguish between these models by varying the sample geometry. However, a simple algebraic exercise shows that the shape drops out and $P/V = E^2/\rho$, so there is no way to distinguish an electric field effect from a power density effect without theoretical guidance. We therefore might consider the hot electron model simply a convenient if accidental analytic description of *R(T,E)* in a range of parameters where (4), for whatever reason, does not apply.

***Time constants and heat capacity.*** However, the hot electron picture has consequences that go beyond a particular *R(T,E)* relation. Since the electron system should have some heat capacity $C_e$, and we have already determined an electron-phonon thermal conductivity $G_{e-ph}$, then there should be a characteristic time $\tau = C_e / G_{e-ph}$ for changes in the electron temperature $T_e$. This has been investigated for NTD Ge [28,29], for ion implanted Si [32,33], and for other disordered systems [34,35]. A thermometer directly attached to the heat sink so that its lattice temperature does not change should look like a simple bolometer as shown in Ch. 1, with the thermal link provided by $G_{e\text{-}ph}$ and the bolometer heat capacity equal to $C_e$. This is clearly observed, and the thermal time constant $\tau$ can be determined quite precisely by measuring the A.C. impedance of the device as a function of frequency and fitting the $Z(\omega)$ function given in Ch 1. Details of this procedure and an example of a measurement of a “tied down” thermometer can be found in [36].

Figure 15 shows $G_{e-ph}(T_e)$ as determined from D.C. I-V curves, $\tau(T_e)$ as determined from A.C. impedance measurements, and their product, which in this hot electron picture should represent the heat capacity of the conduction electron system, $C_e$. Remarkably, while the independently-measured $G_{e-ph}$ and $\tau$ both vary by almost three orders of magnitude over this temperature range, their product is almost constant. Even more remarkable, if one chooses to regard the hot electron model as simply an empirical description of *R(T,E)*, is that a conventional measurement of the total heat capacity of the implanted impurity system, also shown in Fig. 15c, agrees almost perfectly with the derived $C_e$ [33]. This indicates that essentially all of the heat capacity of the impurity system is effectively coupled to the part of the system directly involved in conduction.

The electronic heat capacity, both derived and measured, is considerably flatter than the linear temperature dependence expected for a metallic system. It is almost constant below 0.1 K, then steepens at higher temperatures, approaching $\gamma = 1$ at 0.2 K. This is in good qualitative agreement with other measurements [37,38], and the small shift in absolute value could be due to the difference in compensation (50% vs ~0%). The flat temperature dependence (which becomes negative at sufficiently low doping densities) was predicted and is ascribed to the formation of spin-exchange clusters [39].

***Internal thermodynamic fluctuation noise.*** Another consequence of the literal interpretation of the hot electron model is that additional noise should be observed in the biased thermometer output due to transduced temperature fluctuations of the electron system caused by random energy transport between the electron and phonon systems. Figure 16 shows the measured noise in a “tied down” implant, after the Johnson noise of the resistance is subtracted. The noise and its temperature and bias dependence are readily calculated from the simple bolometer theory outlined in Ch. 1, and the agreement is very good.

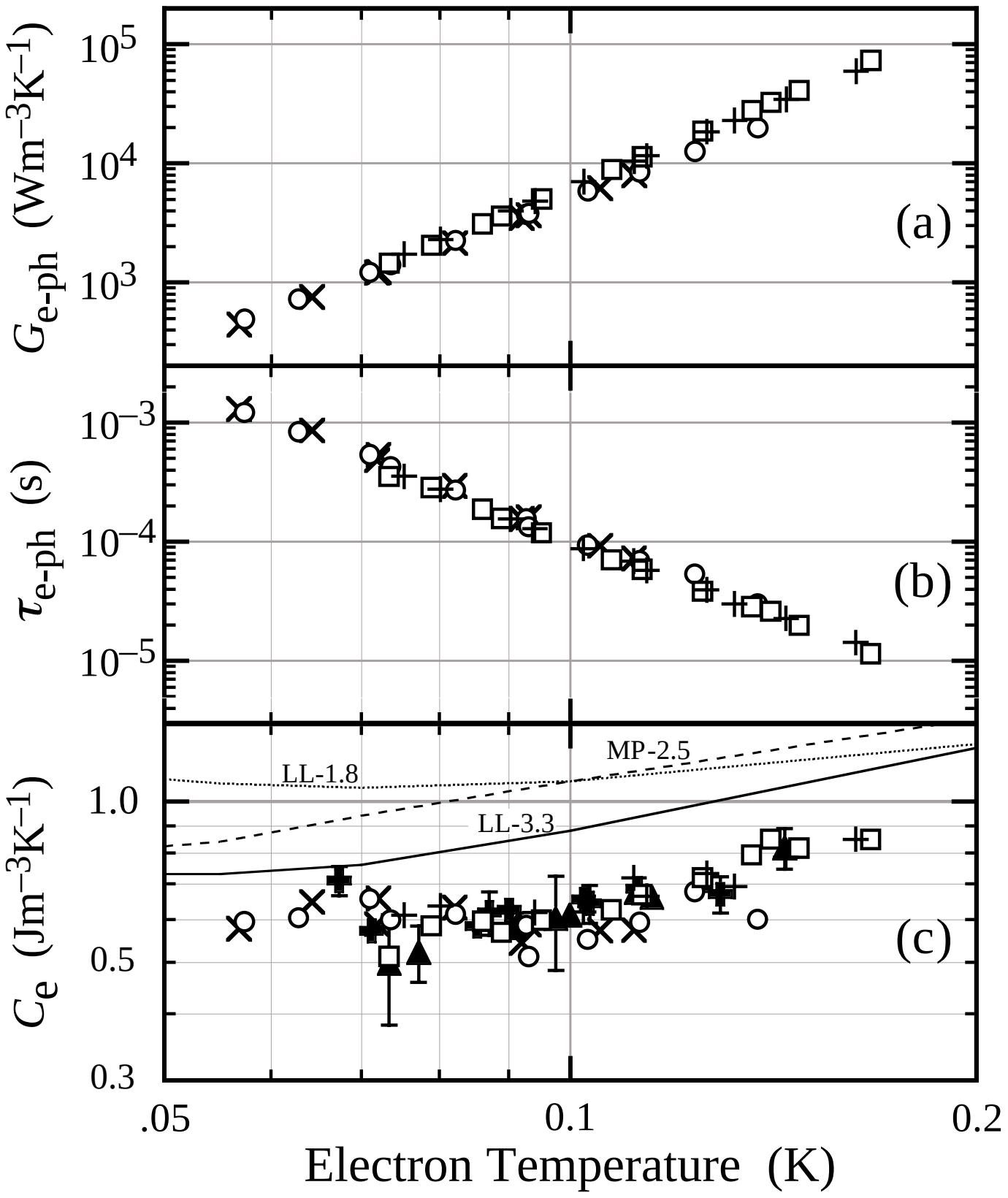

**Fig. 15.** Measurements of ion implanted and diffused Si:P:B devices with net doping densities near $2.5\times10^{18}$ cm$^{-3}$ and 50% compensation ($T_0 \approx 6 - 11$ K). **a)** Coupling constant $G_{\text{e-ph}}$ as a function of $T_\text{e}$ determined from D.C. resistance as a function of bias power. **b)** Characteristic time constant $\tau$ as a function of $T_\text{e}$, determined from A.C. impedance measurements. **c)** Open symbols are the product of a) and b): electronic heat capacity $C_e = G_{\text{e-ph}} \cdot \tau$. The filled symbols are the heat capacity of the implanted silicon measured conventionally by attaching a piece of this material to a thermally isolated platform. The lines show published heat capacity measurements of uncompensated doped Si:P with similar net donor densities from [37] and [38]. (data from [32])

This means that when a doped semiconductor thermometer is used in a bolometer or calorimeter structure with a separate absorber and thermal isolation link, the effective “thermal circuit” must include at least two thermal links and two heat capacities. Internal thermodynamic fluctuations and internal time constants may become important. The simple detector theory of Ch. 1 can be used only if $G_{\text{e-ph}} >> G_{\text{sink}}$ In general, it will be necessary to use more complicated thermal models, such as those derived in [40], that can explicitly include $G_{\text{e-ph}}$ and $C_\text{e}$.

## 4 Excess noise

Additional noise has been observed, primarily in ion-implanted silicon, that appears as an unexpected low-frequency component and has had a significant impact on detector performance. Ion implanted Si detectors made in three different labs according to somewhat different recipes all show similar behavior [41]. Characterization of the noise of these thin (~200 nm) stacked-implant devices over a wide range of doping density,

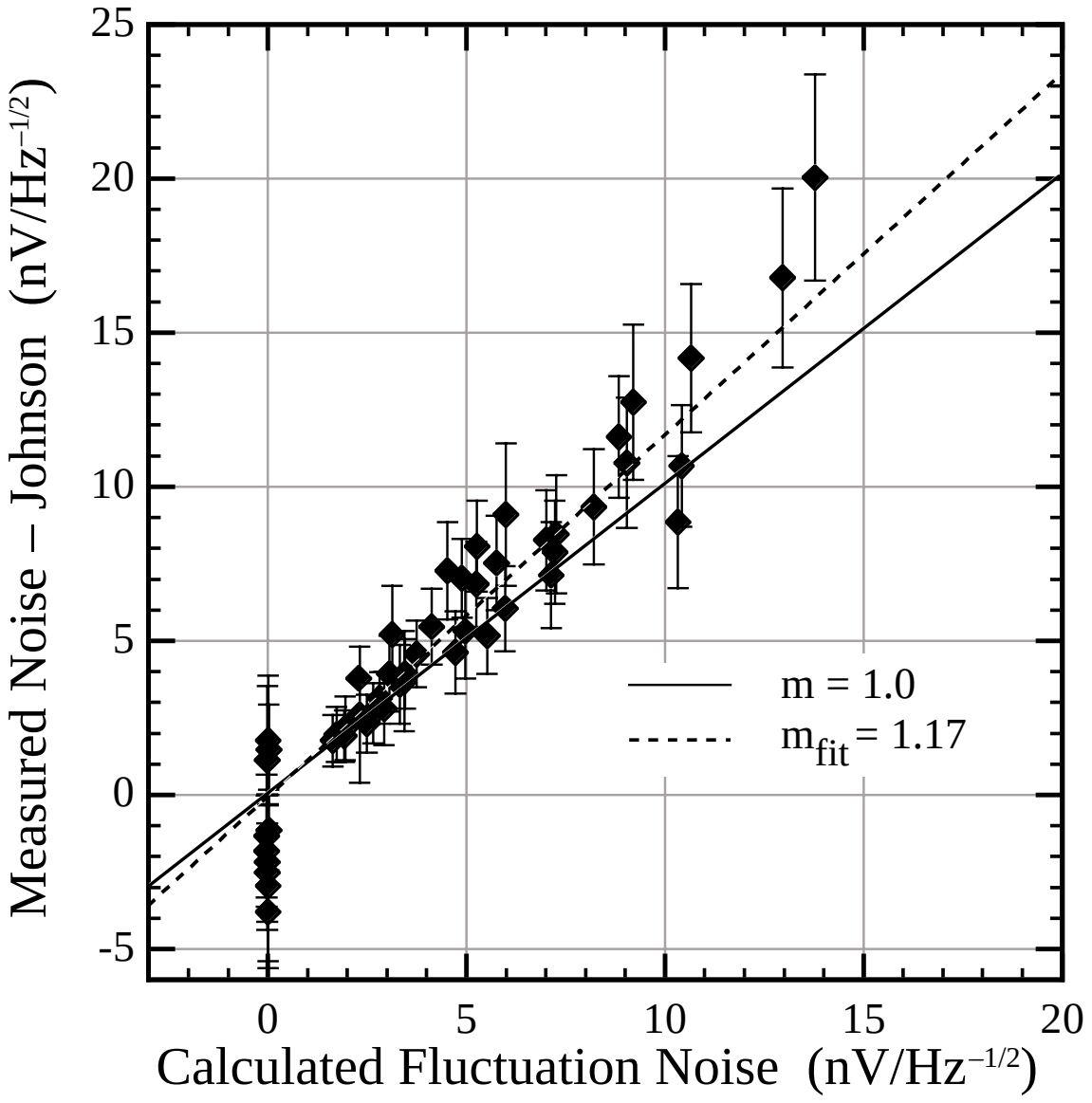

**Fig. 16.** Excess noise observed in an ion implanted Si detector tied down to the heat sink. The observed excess is in reasonable agreement with the predictions of standard bolometer theory for thermodynamic fluctuations over a thermal link with the measured value of $G_{e\text{–}ph}$. (from [32])

lattice temperature, and bias current shows a complicated dependence on all of these factors (including the load resistor value) [42]. When interpreted in terms of the hot electron model described above, however, the behavior after correcting for bias power heating of the electrons and electrothermal feedback effects is greatly simplified and can be characterized as relative resistance fluctuations with a $1/f$ spectral density that depend only on the electron temperature and doping density. Figure 17 shows the resistance fluctuations as a function of $T_e$ for different lattice temperatures and bias currents. Combinations of these that result in the same $T_e$ also give the same derived values for $\left\langle (\Delta R/R)^2 \right\rangle$.

The resistance fluctuations are independent of the shape of the device going from 36:1 to 1:36 length to width ratios, which virtually rules out the ohmic contacts as a significant contributor. They scale quite precisely with (thermistor volume)$^{-1/2}$, as expected for any random fluctuation that is uncorrelated in different parts of the volume. Since the thickness of these devices is limited by the acceleration energy of standard commercial ion implanters, they all have approximately the same thickness. So it is actually only the area$^{-1/2}$ dependence that has been verified. The conventional parameterization for $1/f$ noise is the Hooge-alpha:

$$\left\langle \left( \frac{\Delta R}{R} \right)^2 \right\rangle \equiv \frac{\alpha_{\text{Hooge}}}{Nf} \,, \tag{6}$$

where $N$ is a stand-in for the volume, and is conventionally the number of “carriers”. Han et al [42] arbitrarily use the net number of donors for $N$. Since the doping density

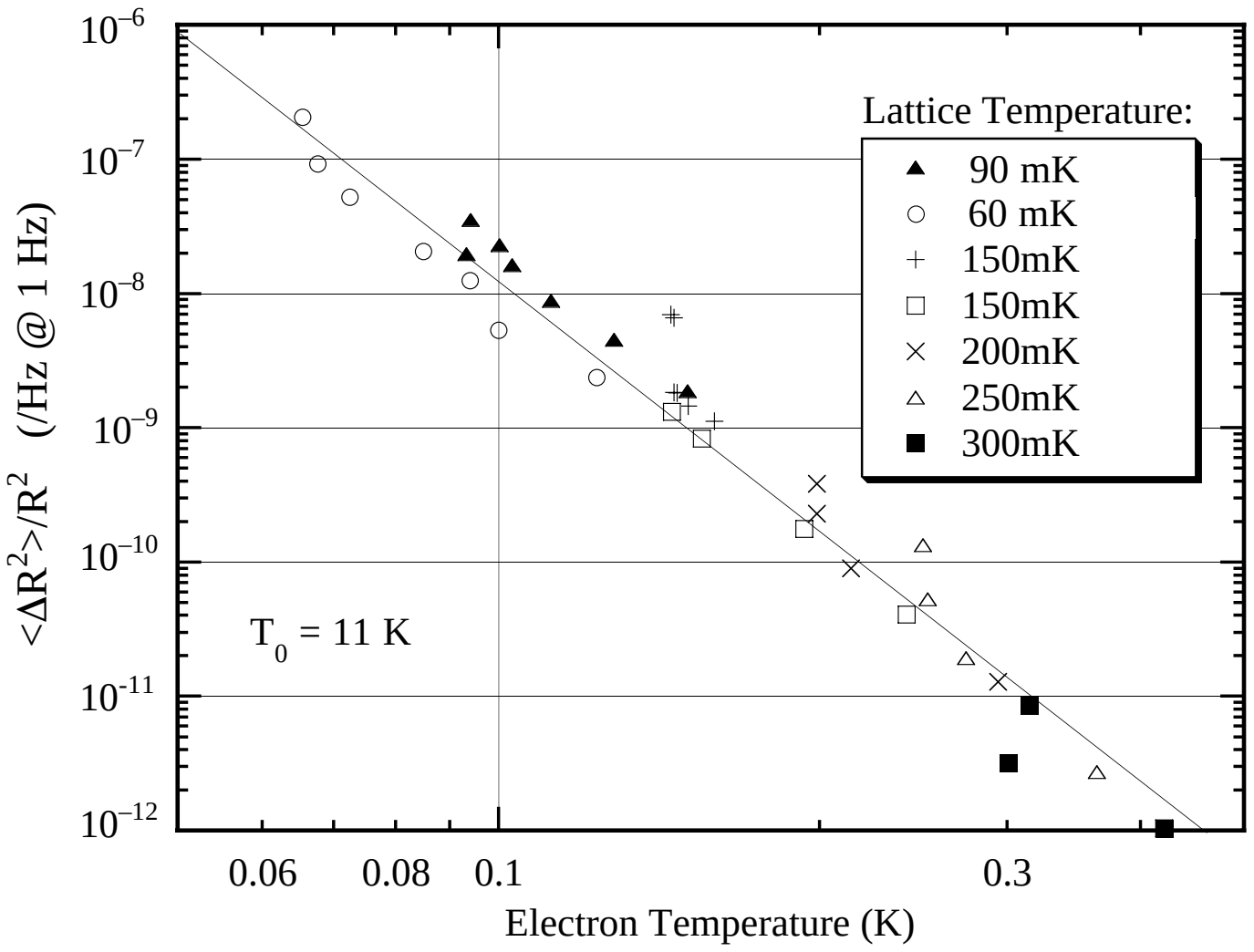

**Fig. 17.** Resistance fluctuations in a thin ion-implanted Si thermometer as a function of the electron temperature $T_e$ for several values of $T_{lattice}$ and bias. The fluctuations appear to depend only on $T_e$, which increases with bias power for a given $T_{lattice}$. (from [42])

does not change very much over the practical range of $T_0$, this is nearly equivalent to volume. Figure 18 shows measurements of $\alpha_{Hooge}$ as a function of $T_e$ for a range of doping densities ($T_0$). These were reasonably well-fit by the empirical function

$$\alpha_{\text{Hooge}} = 0.034 \left(\frac{T_0}{1\text{K}}\right)^{2,453} \cdot \left(\frac{T_e}{0.153\text{K}}\right)^{-(5.2+0.9\log_{10}(T_0/1\text{K}))} . \quad (7)$$

There has been only a little theoretical work on fluctuations in hopping conductivity [43,44], and it is not clear that what has been done is applicable to the usual conditions for low-temperature thermometers. It can be seen from Fig. 18 that the 1/*f* noise increases rapidly as $T_e$ decreases, going approximately as $T_e^{-6}$. Perhaps coincidentally, the size of the percolation network at these temperatures is becoming comparable to the device thickness and is increasing exponentially as the temperature drops, making the conduction rapidly more two-dimensional [6]. Thus the strong temperature dependence offers a hint that the 1/*f* noise is a 2-d or surface effect. NTD Ge thermometers generally do not show significant 1/*f* noise, but most Ge devices are at least 100 µm thick, and there are at least anecdotal reports of it showing up in unusually thin Ge.

The ready availability of good quality silicon-on-insulator wafers with almost any desired device thickness now makes it straightforward to fabricate thicker doped Si thermometers. As described in [20], these have been made by implanting the phosphorus and boron dopants at approximately the middle of a 1.5 µm silicon layer, then annealing at high temperature for a long time so that the impurity atoms diffuse uniformly through the entire thickness. The 1/*f* noise in these devices is reduced by at least a factor of six, as shown in Fig. 19.

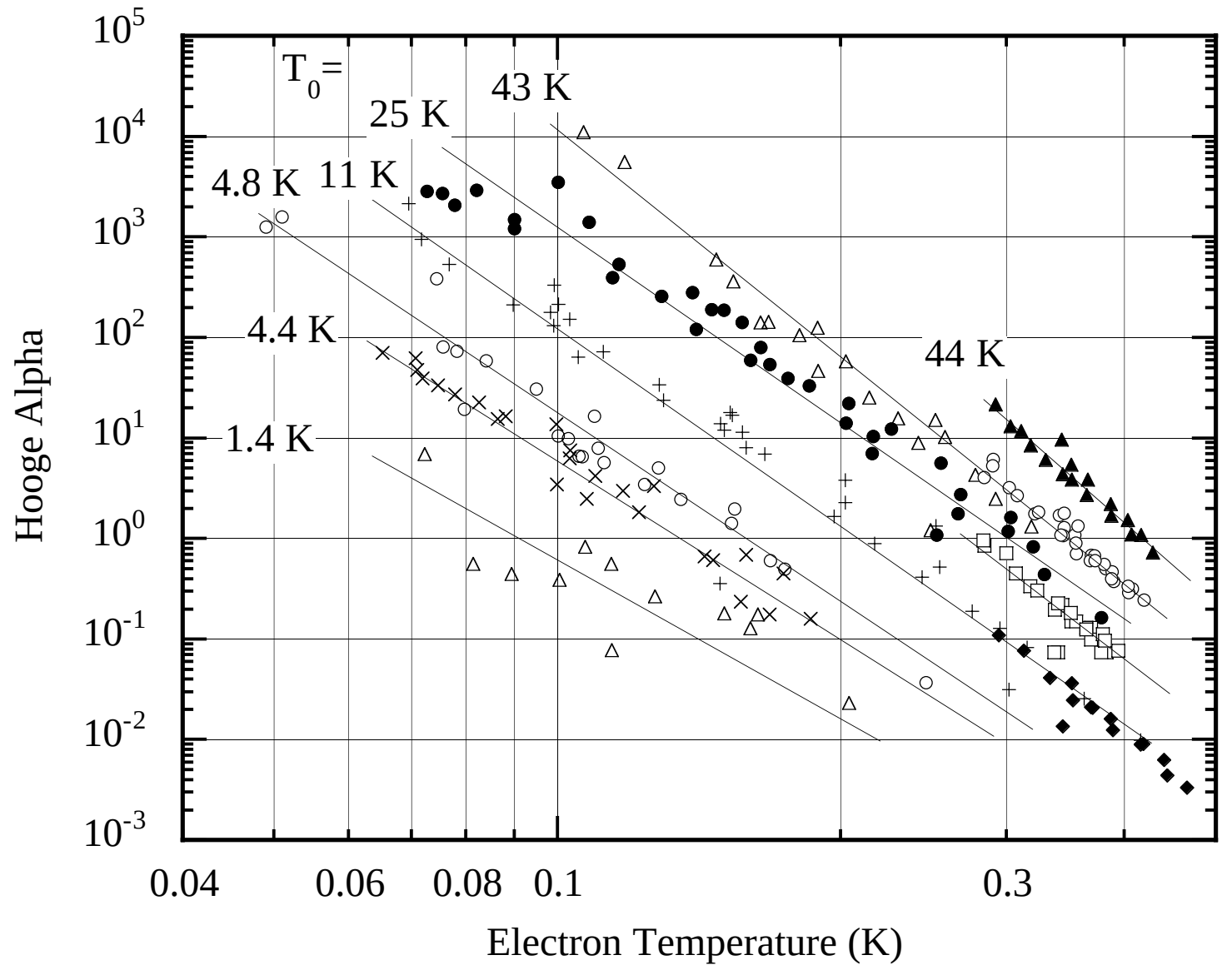

**Fig. 18.** Excess noise parameter $\alpha_{\text{Hooge}}$ for thermistors with different doping densities.

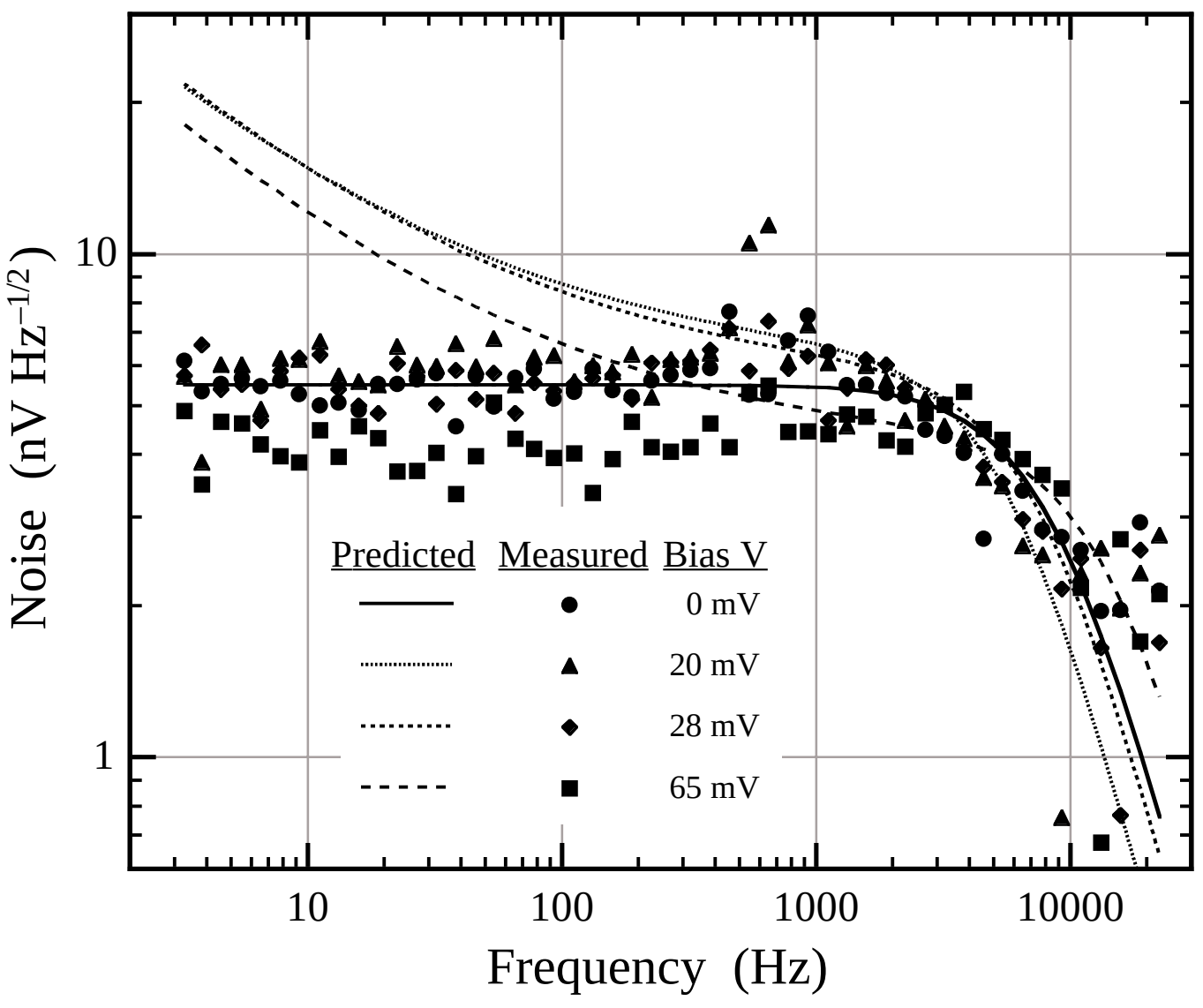

**Fig. 19.** Noise spectra from a 1500 nm thick diffused implant in Si. The lines show the predictions from (7). In the absence of 1/*f* noise, the white noise level falls with increasing bias due to the drop in resistance from bias power heating of the electrons. (from [45])

An alternative explanation for the $1/f$ noise is that while it is a surface effect, it is caused by the proximity of the lightly-doped wings of the stacked-implant profile [46]. Slow tunneling of electrons between sites that are not part of the conduction network could modulate the main conduction path, in analogy to the standard McWhorter theory of low frequency noise in Metal-Oxide-Semiconductor Field Effect Transistors. The improvement is then due less to the increased thickness than to the very abrupt cutoff in the doping density profile. There are obvious experiments that could distinguish between these possibilities, but they have not been done.

## 5 Fundamental Limitations and Ultimate Performance

The "non-ideal" effects described above all limit the performance of detectors with semiconductor thermometers. The linear analysis of Ch. 1 gives the logarithmic sensitivity $\alpha$ as the only significant thermometer parameter. Hot electron and field effects limit practical values of $\alpha$ to 10 or less in most cases — far less than the sensitivity currently available with superconducting transition edge thermistors. So why use semiconductors at all? The answer to this is certainly application-dependent. The most obvious consideration is large signals, where the saturation characteristics of transition edge sensors are best described as "less than graceful." Semiconductor thermometers become more nonlinear and less sensitive with increasing signal size, but have no hard saturation limits. They are easier to characterize and can also be easier to fabricate and simpler to use. The ideal amplifier for semiconductor thermometers is a simple Junction Field Effect Transistor (JFET) costing less than $1. On the other hand, for large numbers of detectors practical advantage can swing strongly in favor of the SQUID amplifiers used with superconducting thermometers.

One perhaps fundamental advantage of semiconductors is that they are relatively insensitive to magnetic fields. They have been operated without degradation in fields as large as 10 T [47]. The characteristics shift considerably, but the thermometer can be optimized for a given field. Conversely, a single thermometer can be made to perform over a wide range of temperatures by "tuning" it with a variable magnetic field.

Even the shortcomings can occasionally be put to good use. The hot electron effect allows a simple thermometer to be used as a complete detector, with an absolute minimum of addenda. The electron system is both the absorber and the thermometer, and the thermal isolation suspension is provided by the electron-phonon coupling. The usual drawback of these hot electron bolometers is that they have few adjustable parameters and cannot be optimized for a particular situation. However, the built-in characteristics of thin silicon implants are almost ideal for the requirements of some infrared detector applications [48,49]. The R.F. sheet resistance is a good match to free space or a simple transmission line, so the absorber is highly efficient. Time constants cover a reasonable range at temperatures where the thermodynamic performance is good, and the low-frequency resistance for a square detector is a good match to the noise resistance of a JFET amplifier. It appears that very simple detectors could be made with outstanding performance.

## 5.1 Optimization

Getting the best possible performance requires optimizing the thermometer design. To do this, one needs both an understanding of how the thermometer works and quantitative data on the behavior of key parameters. We now seem to have an adequate phenomenological understanding of how semiconductor thermometers work, even if the underlying theory is a little murky. The hot electron model should be taken at face value and modeled as a separate thermal conductance and heat capacity. This complicates the thermal circuit (aside from the happy case of the hot electron bolometer) but appears to account correctly for the additional time constants and noise sources that are observed, and formalisms exist for handling arbitrary thermal models [40,50]. The field effect can be included simultaneously. In the linearized performance theory of Ch. 1 it shows up as the parameter $\beta \equiv \partial \log R / \partial \log V)_T$. For the field effect model (4), $\beta$ is simply $-CeE\lambda / kT$. Of course, to perform the optimization, one has to know how $\beta$, $G_{e\text{-}ph}$, and $C_e$ depend on other parameters.

We are not so well off with the necessary engineering data. The data for silicon are barely adequate for low $T_0$ material operated at not too low temperatures where electric field effects are negligible. For NTD germanium both field effect and electron heating are usually important, and there is a surprising scarcity data on either. Part of the problem is that little analysis has been done that takes both effects into account, and this has confused the results. Now that the situation is better understood, it is relatively easy to separate parameters and evaluate both effects simultaneously. It is straightforward to measure the A.C. impedance $Z(\omega)$, and since $\beta = 1 - R/Z(\infty)$, $\beta$ and with it the local value of $\lambda$ can be extracted immediately from its high frequency limit. This can be used to separate heating and field effects at the bias point, giving a value for $G_{e\text{-}ph}$, which along with $\beta$ can then be put into the fit of $Z(\omega)$ to get $\tau$ and $C_e$.

## 5.2 Germanium vs Silicon

The choice of Si or Ge depends very much on the application, and is usually decided by non-performance characteristics such as the excellent reproducibility of NTD Ge or the ease of fabricating large numbers of small integrated thermometers with ion-implanted Si. However, it is still interesting to consider whether there is some universal figure of merit that favors one or the other. The beneficial thermometer parameter is its sensitivity, while its heat capacity has a negative impact on detector performance. Both electric field and hot electron effects reduce the effective sensitivity as the power density goes up and a rough figure of merit would be sensitivity at a given power per unit heat capacity and temperature. This is a complex quantity, since field effect and electron heating affect the sensitivity in different ways.

Field effect (non-zero $\beta$) reduces both signal and thermodynamic fluctuation noise by the same amount, so if amplifier noise is not significant, the only change in signal to noise ratio comes from Johnson noise. Johnson noise in biased nonlinear resistances is a diffiicult theoretical subject [51,52] that we will avoid here, since hot electrons generally have more impact on detector performance than field effect. (The opposite is the case for superconducting transition edge thermometers, so see the next chapter for more discussion.) In the usual case where the signal comes in through the phonon system, the sensitivity to phonon temperature is rapidly degraded by electron heating. This can be seen from (5), remembering that the exponent is ~6. The thermodynamic fluctuations between the electron and phonon baths can create additional noise, and the $C_e/G_{\mathrm{e-ph}}$ time constant rolls off the signal spectrum with respect to the thermometer Johnson noise and will degrade the overall signal to noise ratio unless it is shorter than the primary thermal time constant of the detector by a factor of ~α. For all of these hot electron effects, the figure of merit is $G_{\mathrm{e-ph}}/C_e$ at a given thermometer sensitivity $\alpha \approx 0.5(T_0/T)^{1/2}$. This quantity is independent of thermometer volume, and is plotted in Fig. 20 as $\tau_{\mathrm{e-ph}} \equiv C_e/G_{\mathrm{e-ph}}$. Data for NTD Ge are very sparse and inconsistent by a factor of ten or more. On the average, however, it appears that the intrinsic time constants may be faster by a factor of about six for the same $T_0$ and $T_{\mathrm{e}}$. Much better data are needed for design optimization and predictions of ultimate performance, and it should be straightforward to make the necessary measurements.

### 5.3 Expected performance

We are at a point where performance can be predicted with reasonable accuracy when thermal data are available. A 32-detector array, identical to the XRS-2 array shown in Fig. 8, was built for an atomic physics experiment and operated at 60 mK. It had measured resolution within 10% of the calculated 4.8 eV F.W.H.M. for 6 keV X-rays on (almost) all of the detectors. Given the particular X-ray absorber used, the main avenue for improving the resolution would be to improve the thermal coupling between the absorber and thermistors. With perfect coupling, the predicted resolution would be 3.4 eV, but problems with “thermalization noise”, where variable numbers of super-thermal phonons reaching the thermometer produce pulse-to-pulse variations in the response, would actually make the resolution worse. The optimum solution for this is unknown, and the phonon physics is complex, so progress requires experimenting with different absorber couplings. Lowering the operating temperature could improve the resolution further, and germanium thermometers might be better, but we don't have the data to say by how much or under what conditions.

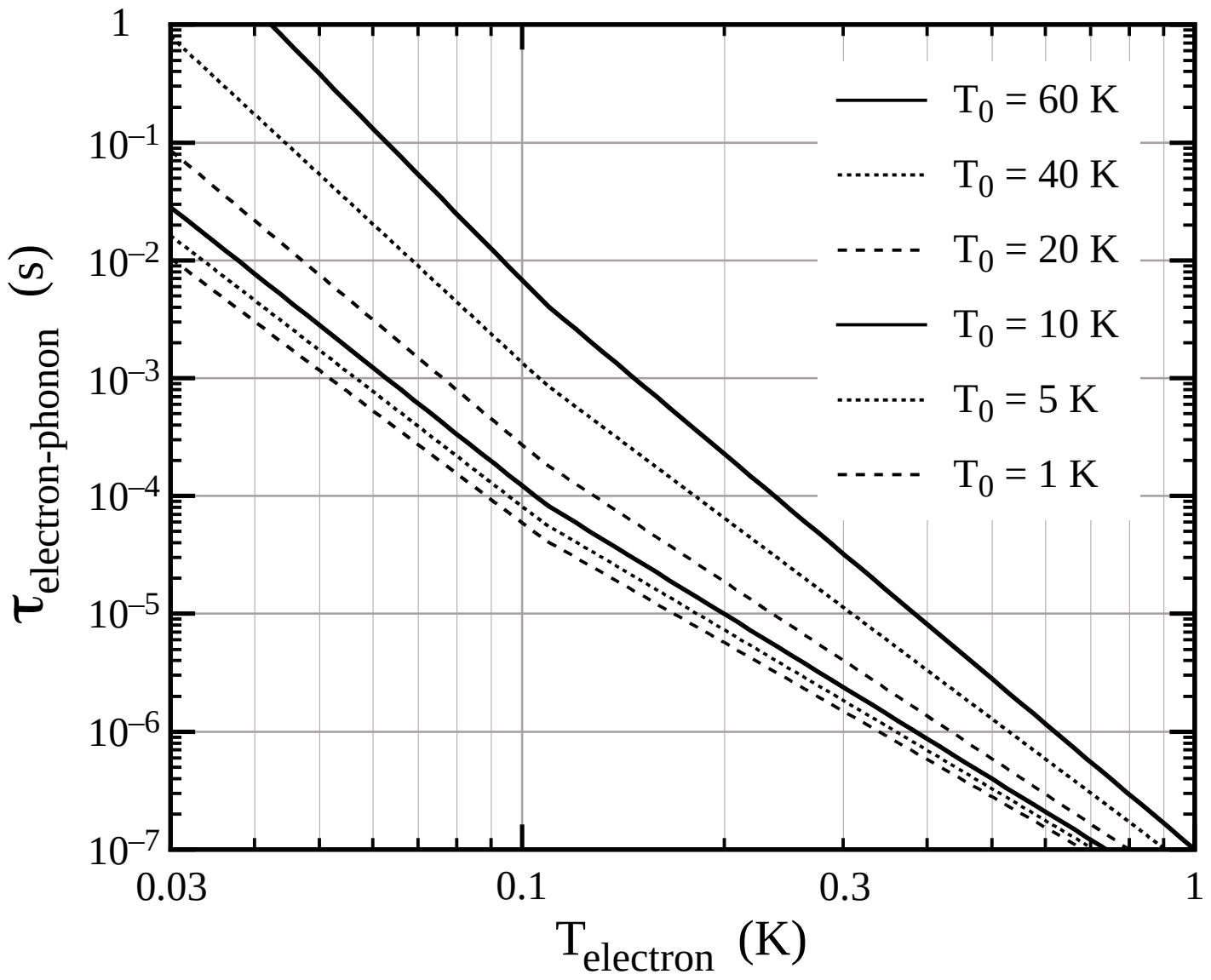

**Fig. 20.** The electron-phonon time constant $C_e/G_{\mathrm{e-ph}}$ vs electron temperature. The solid lines are for Si with doping density given by the indicated $T_0$ using fits to $G_{\mathrm{e-ph}}$ data from [24] with normalization and $C_e$ from [33]

One of the better X-ray detector results obtained so far with a semiconductor thermometer is shown in Fig. 21. There is some room for improvement: one might expect that if everything could be optimized, a 2 eV detector with these parameters is feasible. But for a detector of this heat capacity operating at ~50 mK, some other thermometer technology will be necessary to reach 1 eV.

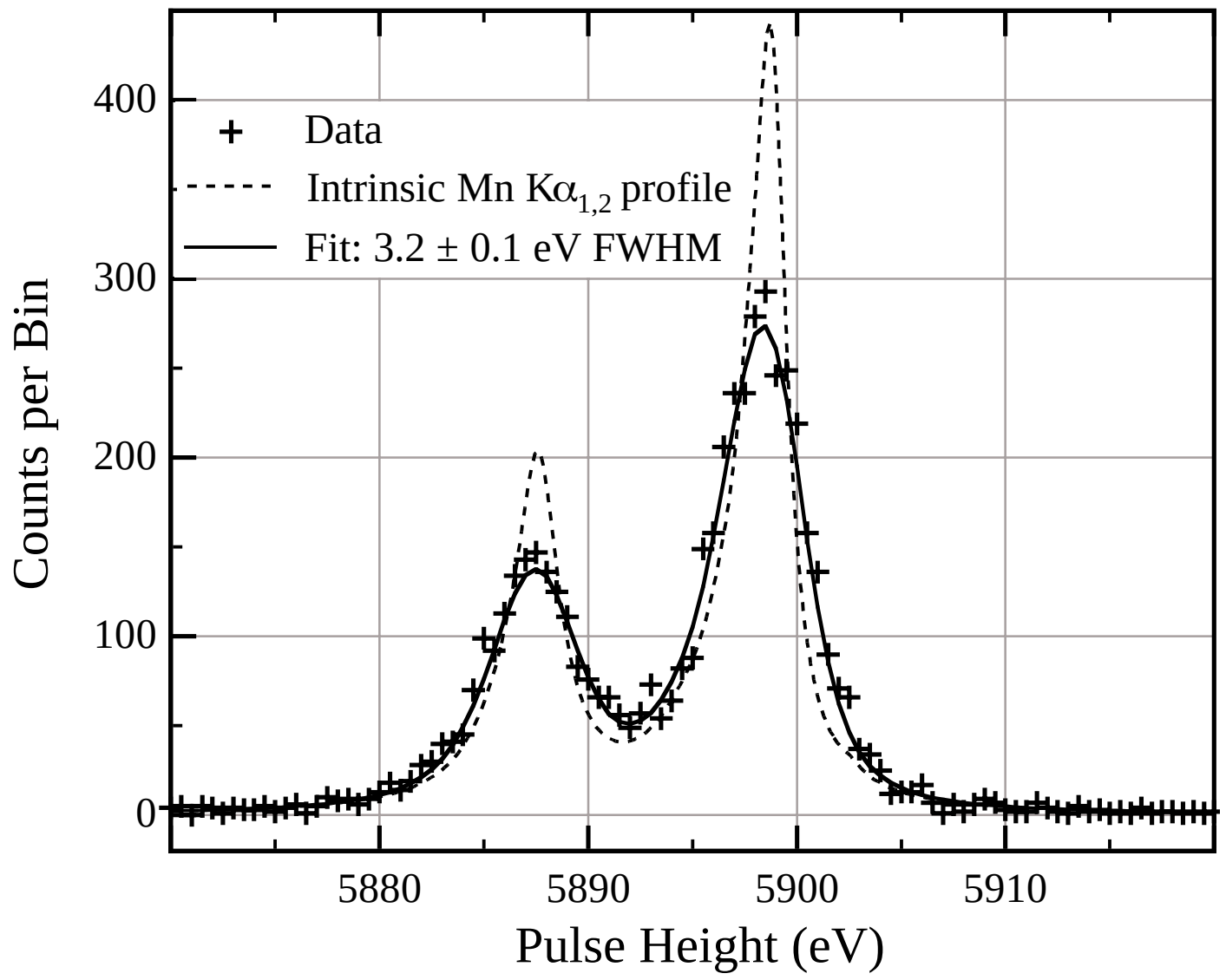

**Fig. 21.** Pulse height spectrum of the $^{55}$Mn $K\alpha_1$ and $K\alpha_2$ lines at 5.9 keV, showing a resolution of 3.2 ± 0.1 eV FWHM. This device had a uniformly illuminated 0.41 x 0.41 $mm^2$ HgTe absorber 8 µm thick, which provides better than 95% X-ray stopping efficiency up to 7 keV and 50% up to 17 keV. It was operated at a detector temperature of ~60 mK

I would like to thank E.E. Haller and J.W. Beeman for providing their data on NTD Ge prior to publication. Many members of the Wisconsin and NASA/Goddard X-ray groups contributed heavily to this chapter. I thank in particular Jiahong Zhang, Mike Juda, Massimiliano Galleazzi, Andy Szymkowiak, Scott Porter, Caroline Kilbourne, Richard Kelley, and Regis Brekosky. This work was supported in part by NASA grant NAG5-5404.

## References

1. R. C. Jones, J. Opt. Soc. Am. **37**, 879 (1947)
2. I. Esterman, Phys. Rev. **78**, 83 (1950)
3. F. J. Low, J. Opt. Soc. Am. **51**, 1300 (1961)
4. P.W. Anderson, phys. Rev. **109**, 1492 (1958)
5. N.F. Mott & W.D. Twose, Adv. Phys. **10**, 107 (1961)
6. B.I. Shklovskii & A.L. Efros, Electronic Properties of Doped Semiconductors (Springer, Berlin 1984)
7. A.L. Efros & B.I. Shklovskii, J. Phys. C **8**, L49 (1975)
8. J. Zhang *et al.*, Phys. Rev. B **48**, 2312 (1993)
9. J. Zhang *et al.*, Phys. Rev. B **57**, 4950 (1998; erratum to [8])
10. I.S. Shlimak, in *Hopping and Related Phenomena* eds. H. Fritzsche & M. Pollak (World Scientific, Singapore 1990) p. 49
11. P. Dai, Y. Zhang, & M.P. Sarachik, Phys. Rev. Lett. **69**, 1804 (1992)
12. W. Bergmann-Tiest, personal communication (1998)
13. L.R. Semo-Scharfman, personal communication (2003)
14. A.L. Woodcraft *et al.*, J. Low Temp. Phys. **134**, 925 (2004)
15. E.E. Haller, N.P. Palaio, M. Rodder, W.L. Hansen, and E. Kreysa, in *Neutron Transmutation Doping of Semiconductor Materials*, ed. R.D. Larrabee (Plenum, New York 1984) p 21
16. Itoh *et al.*, J. Low Temp. Phys. 93, 307 (1993)
17. E. E. Haller & J.W. Beeman, personal communication (2004)
18. C. Arnaboldi *et al.*, Nucl. Instr. Methods **A518**, 775 (2004)
19. A.D. Turner *et al.*, Appl. Optics **40**, 4921 (2001)
20. R.P. Brekosky *et al.*, Nucl. Instr. & Methods **A520**, 439 (2004)
21. P.M. Downey *et al.*, Appl. Optics **23**, 910 (1984)
22, S.H. Moseley, J.C. Mather, & D. McCammon, J. Appl. Phys. **56**, 1257 (1984)
23. R.M. Hill, Philos. Mag. *24*, 1307 (1971)
24. J. Zhang *et al.*, Phys. Rev. B **57**, 4472 (1998)
25. T.W. Kenny *et al.*, Phys. Rev. B **39**, 8476 (1989)
26. S.M. Grannan, A.E. Lange, E.E. Haller, and J.W. Beeman, Phys. Rev. B **45**, 4516 (1992)
27. M. Piat *et al.*, J. Low Temp. Phys. **125**, 189 (2001)
28. N. Wang *et al.*, Phys. Rev. B **41**, 3761 (1990)
29. E. Auberg *et al.*, J. Low Temp. Phys. **93**, 289 (1993)
30. B.I. Shklovskii, personal communication (2004)
31. M. Prunnila *et al*, Physica E **13**, 773 (2002)
32. D. Liu et al., in *Low Temperature Detectors*, (F.S. Porter et al. Eds.), Proc. 9th Int'l Workshop on Low Temperature Detectors in Madison, Wisconsin, July 2001 (AIP, New York 2002) p 87. (see also M. Galeazzi et al., Nucl. Instr. Methods A **420**, 469 (2004))
33. M. Galeazzi et al., Phys. Rev. B (submitted)
34. M.A. LaMadrid, W. Contrata, J.M. Mochel, Phys. Rev. B **45**, 3870 (1992)
35. S. Marnieros, L. Bergé, A. Juillard, & L. Dumoulin, Phys. Rev. Lett. **84**, 2469 (2000)

36. J.E. Vaillancourt, Rev. Sci. Instr. (in press)
37. M.A. Paalanen, J.E. Graebner, R.N. Bhatt, Phys. Rev. Lett. **61**, 597 (1988)
38. M. Lakner, H.v. Löhneysen, Phys. Rev. Lett. **63**, 648 (1989)
39. R.N. Bhatt, P.A. Lee, Phys. Rev. Lett. **48**, 344 (1982)
40. M. Galeazzi, D. McCammon, J. Appl. Phys. **93**, 4856 (2003)
41. A. Nucciotti, personal communication (1998)
42. S-I Han *et al.*, in *EUV, X-Ray and Gamma-Ray Instrumentation for Astronomy IX*, (O.H. Siegmund, M.A. Gummin, Eds.), Proc. SPIE v.**3445**, 640 (1998)
43. B.I. Shklovskii, Solid State Commun. **33**, 273 (1980)
44. Sh.M. Kogan, B.I. Shklovskii, Fiz. Tekh. Poluprovodn. **15**, 1049 (1981) [Sov. Phys. Semicond. **15**, 605 (1981)]
45. D. McCammon *et al*., in *Low Temperature Detectors*, (F.S. Porter et al. Eds.), Proc. 9th Int'l Workshop on Low Temperature Detectors in Madison, Wisconsin, July 2001 (AIP, New York 2002) p 91
46. S.H. Moseley, personal communication (2001)
47. P. De Moor *et al.*, J Low Temp. Phys. **93**, 295 (1993)
48. H. Moseley, D. McCammon, in *Low Temperature Detectors*, (F.S. Porter et al. Eds.), Proc. 9th Int'l Workshop on Low Temperature Detectors in Madison, Wisconsin, July 2001 (AIP, New York 2002) p 103
49. T.R. Stevenson et al., in *Millimeter and Submillimeter Detectors for Astronomy II*, (J. Zmuidzinas, W.S. Holland Eds.), Proc. SPIE v.**5498** (in press)
50. E. Figueroa-Feliciano, Ph.D. thesis, Stanford University (2001)
51. H.B. Callen & T.A. Welton, Phys. Rev. **83**, 34 (1951)
52. L. Weiss & W. Mathis, IEEE Electron Device Lett. **20**, 402 (1999)